\documentclass{article}

\usepackage[preprint]{neurips_2025} 
\usepackage{ulem}
\usepackage{float}
\usepackage{wrapfig}
\usepackage{array}
\usepackage[utf8]{inputenc}
\usepackage{listings}

\workshoptitle{Multi-Turn Interactions in Large Language Models}

\usepackage[utf8]{inputenc} 
\usepackage[T1]{fontenc}    
\usepackage{hyperref}       
\usepackage{enumitem}       
\usepackage{url}            
\usepackage{booktabs}       
\usepackage{amsfonts}       
\usepackage{nicefrac}       
\usepackage{microtype}      
\usepackage{xcolor}         
\usepackage[pdftex]{graphicx} 
\usepackage{amsmath, amssymb}
\usepackage{algorithm}
\usepackage[noend]{algpseudocode}
\usepackage{subcaption} 
\usepackage{array}
\usepackage[most]{tcolorbox} 
\usepackage{booktabs} 
\usepackage{tabularx}
\usepackage{subcaption}
\usepackage{listings}
\usepackage{xcolor}

\lstdefinestyle{prompt}{
  basicstyle=\ttfamily\small,
  breaklines=true,
  frame=single,
  backgroundcolor=\color{gray!5},
  columns=fullflexible
}

\title{Orchestrator: Active Inference for Multi-Agent Systems in Long-Horizon Tasks}
\workshoptitle{Orchestrator: Active Inference for Multi-Agent Systems in Long-Horizon Tasks}

\author{
  Lukas Beckenbauer$^{*\dagger}$\thanks{Corresponding author} \and
  \textbf{Johannes Löwe}$^{\dagger}$ \and
  \textbf{Ge Zheng}$^{*}$ \and
  \textbf{Alexandra Brintrup}$^{*}$ \\
  $^*$Department of Engineering, University of Cambridge, Cambridge, UK \\
  $^\dagger$TUM School of Management, Technical University of Munich, Munich, DE
}

\begin{document}

\maketitle

\begin{abstract}
Complex, non-linear tasks challenge LLM-enhanced multi-agent systems (MAS) due to partial observability and suboptimal coordination. We propose Orchestrator, a novel MAS framework that leverages attention-inspired self-emergent coordination and reflective benchmarking to optimize global task performance. Orchestrator introduces a monitoring mechanism to track agent-environment dynamics, using active inference benchmarks to optimize system behavior. By tracking agent-to-agent and agent-to-environment interaction, Orchestrator mitigates the effects of partial observability and enables agents to approximate global task solutions more efficiently. We evaluate the framework on a series of maze puzzles of increasing complexity, demonstrating its effectiveness in enhancing coordination and performance in dynamic, non-linear environments with long-horizon objectives.
\end{abstract}
\section {Introduction}
With the rapid advancement of Large Language Models (LLMs), research on intelligent multi-agent systems (MAS) is gaining new traction. Researchers have investigated use-cases across a broad range of applications, including enhancing the reasoning and task-execution capabilities of general-purpose LLMs \citep{chen_agentverse_2024, li_more_2024, zhang_exploring_2024}, supporting software production in recommender systems \citep{liu_lessons_2025, he_llm-based_2025}, facilitating data interfacing and visualization \citep{Xu_vis-analysis_2025}, and enabling self-supervising supply chain infrastructures \citep{xu_multi-agent_2024, xu_implementing_2024}.  However, while a need for AI-driven MAS solutions that enable advanced agent-coordination and effectiveness across complex, non-linear task-settings has been recognized \citep{alon_multiagent_2020, chen_agentverse_2024, chen_optima_2025, li_more_2024, yao_hdflow_2024, prakki_active_2025}, research on the optimization of system-level MAS-coordination towards task execution for \textit{non-linear}, \textit{long-horizon} problem settings has gained traction only recently \citep{xu_multi-agent_2024}.

Existing work has primarily advanced MAS by improving feedback loops across agent-to-agent or agent-to-environment settings. In traditional settings this has mainly been pursued via reinforcement learning \citep{iqbal_actor-attention-critic_2018, liu_grounded_2024, ding_multi-agent_2024, erdogan_plan-and-act_2025}. More recently, attention has shifted toward LLM-supported multi-agent collaboration, often enhanced by reflective or supervisory mechanisms \citep{zhuge_language_2024, nayak_long-horizon_2024, chang_sagallm_2025, erdogan_plan-and-act_2025}. While these approaches have demonstrated notable success, they often rely on static topologies \citep{chen_agentverse_2024, chen_optima_2025, li_more_2024, li_adaptive_2025} and are typically benchmarked on short-horizon, agent-specific tasks such as HumanEval, GPQA \citep{rein2024gpqa}, or GSM8K \citep{kapoor_ai_2025}. Further, while a dominant body of this work has focused on improving planning efficiency in long-horizon settings of steady levels of complexity \citep{dang_multi-agent_2025, zhang_optimizing_2025, erdogan_plan-and-act_2025, xiao_tradingagents_2025, walters_fe_risks_2025}, there remains limited exploration into how LLM-augmented MAS can be enabled to scale and sustain high-accuracy when addressing advanced, long-horizon tasks characterized by growing complexity levels \citep{shojaee_illusion_2025, dao2025alphamaze}.

To address this challenge, we propose \textbf{Orchestrator}---a multi-agent coordination framework with task-observation instance and embedded active inference feedback-loops---and apply it to solving a series of classic maze puzzles with varying difficulty levels (easy, medium, hard). Grounded in active inference principles \citep{parr_active_inference_2022, walters_fe_risks_2025}, stating that sentient agents act to minimize surprise and maintain their internal states by minimizing a quantity called variational free energy (VFE), Orchestrator draws on a benchmark-driven introspection mechanism that considers both, inter-agentic communication \citep{bo_reflective_2025, erdogan_plan-and-act_2025, prakki_active_2025}, and dynamic states between agents’ and their immediate environment \citep{yao_hdflow_2024, bo_reflective_2025, ruiz-serra_factorised_2025}. We operationalize active inference by contrasting agent's realized information gain with coordination costs and optimizing for free energy (FE) output as a measure of effective task solving. This signal regulates agent autonomy and dynamically adapts system behavior in response to rising decision uncertainty and/or efficiency costs \citep{suri_surprise_2022, ruiz-serra_factorised_2025, yeganeh_deep_2025}. Subsequently, we address agents’ partial observability, as a key limitation to overall operational performance \citep{omidshafiei_deep_2017}, by formulating the iterative approximation of effective agent-to-agent and agent-environment coordination as a quantitative optimization problem.

We evaluate Orchestrator's capacity to overcome local minima, by testing it on a range of maze puzzles with varying levels of escape complexity. 
Orchestrator outperforms baseline agent ensembles that operate without active-inference benchmarking and dynamic orchestration by an average factor of 3,03 on mazes of 18 x 18 size and medium difficulty. Specifically, our results indicate that, compared to a baseline success rate of 11\%, active inference-driven orchestration significantly improves reliability, efficiency, and scalability in long-horizon maze-solving tasks, achieving up to 100\% accuracy across 25 runs in medium difficulty and up to 76,67\% accuracy in hard mazes of 25 x 25 size.

We validate our results through ablation studies and summarize our key contributions as: (i) we introduce a self-optimizing, scalable cell architecture consisting of a planning, execution, and observation instance—driven by active-inference feedback and task-observation mechanisms, enabling MAS to operate effectively in settings that demand adaptive autonomy; (ii) we propose a set of coordination benchmarks and optimization methods that track both agent's internal decision outcomes and their collaborative behavior, guiding agent cells to away from local task-completion minima and toward globally optimal solution horizons; and (iii) we demonstrate sustained task-completion accuracy across long-horizon maze tasks across various difficulty levels, using only lightweight, resource-efficient LLM models, thus aligning with production-ready deployment scenarios under strict budget and resource constraints.

\begin{figure}[t!]
  \centering
  \begin{subfigure}[b]{0.30\linewidth}
  \vspace{-5mm}
    \centering
    \includegraphics[width=\linewidth]{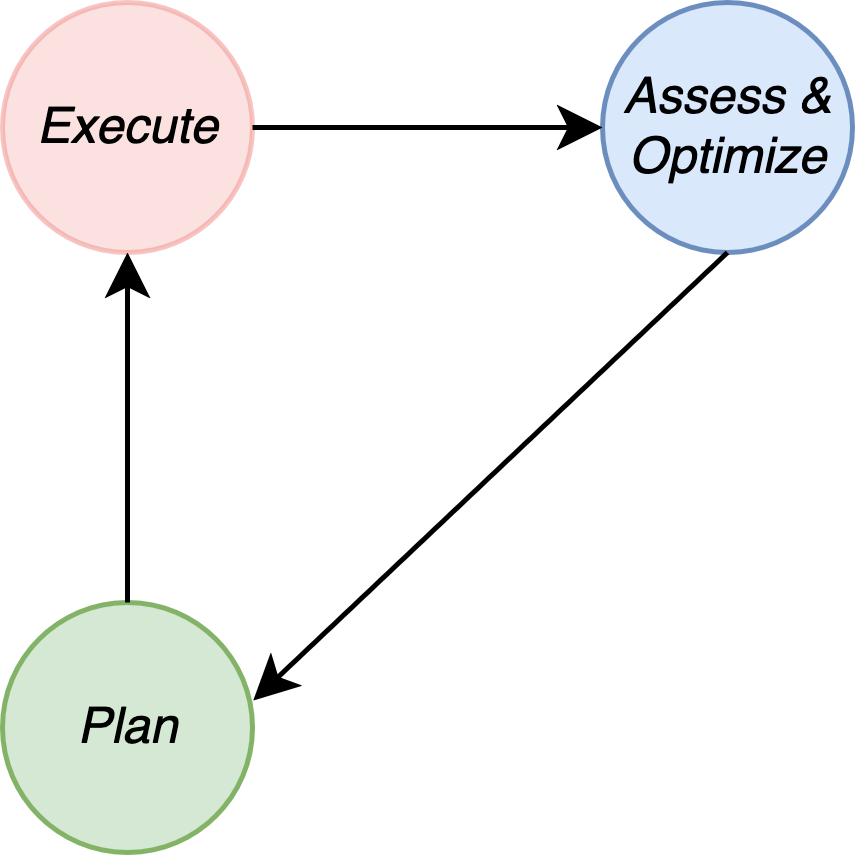}
    \caption{Graph Depiction of Cell-Internal MAS Workflow}
    \label{fig:orch-sub1}
  \end{subfigure}
  \hfill
  \begin{subfigure}[b]{0.68\linewidth}
    \centering
    \includegraphics[width=\linewidth]{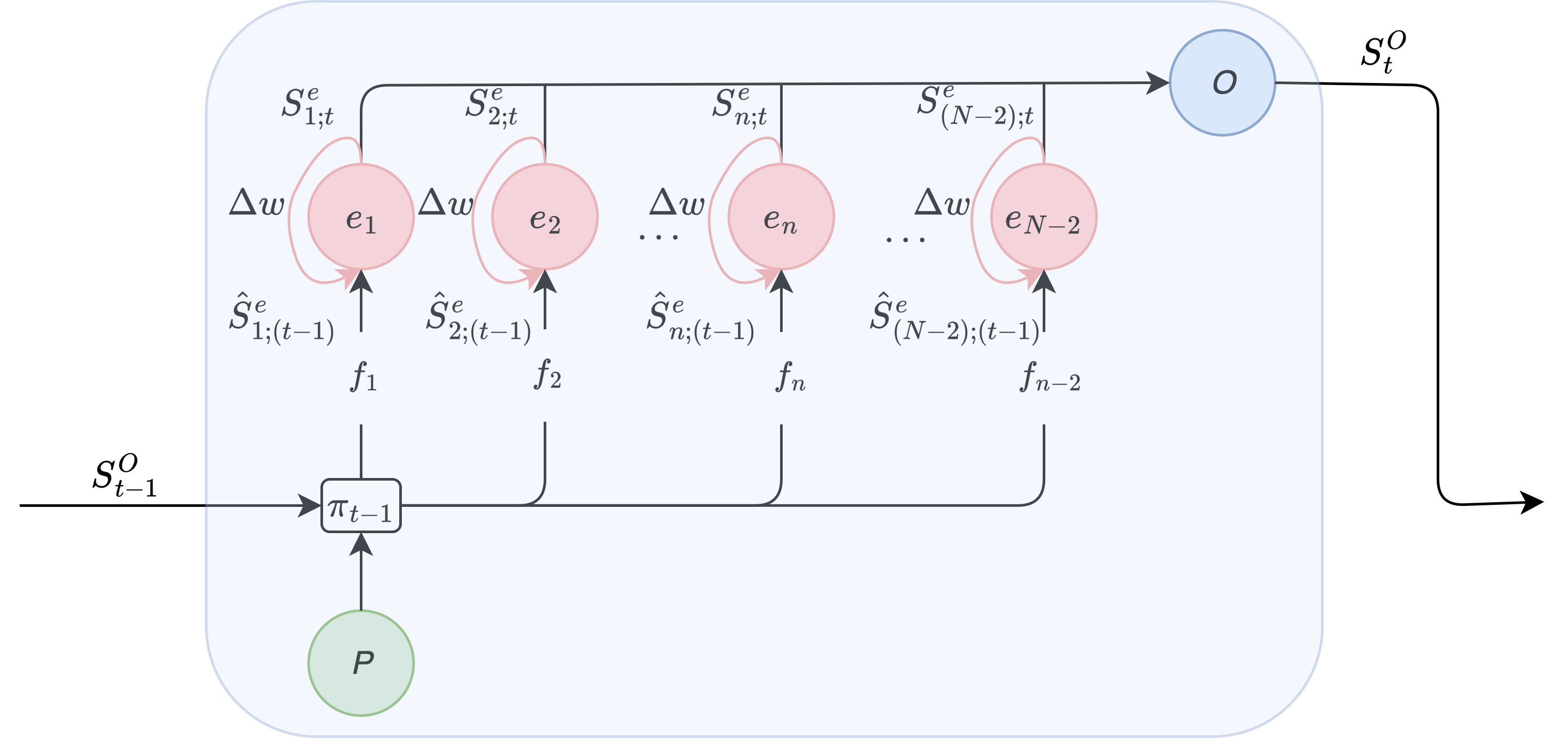}
    \caption{Orchestrator Cell Design. A reprint is available in Appendix \ref{appendix:orchestrator-framework_large}.}
    \label{fig:orch-sub2}
  \end{subfigure}
  
  \caption{Orchestrator Framework Overview}
  \label{fig:orchestrator-framework}
  \vspace{-5mm}
\end{figure}

\section{Related Work}
\label{related_work}
\paragraph{Maze-Assessment as Long-Horizon Benchmark for MAS.} Maze-based environments have become a central testbed for evaluating the reasoning, planning, and coordination abilities of intelligent agents \citep{linardakis_distributed_2024, dao2025alphamaze, godin2025amaze}. Early benchmarks focus on single-agent navigation in static, fully observable mazes, where classic algorithms such as A*, Flood Fill, DFS, or multi-agent pathfinding (MAPF) \citep{stern2019multi} in non-LLM contexts are used to assess pathfinding and basic spatial reasoning capabilities \citep{foead_systematic_2021, tjiharjadi_systematic_2022, liu_cooperative_2025, godin2025amaze}. 

More recent work has produced a new generation of maze benchmarks that probe the limits of agent memory, adaptability, and sequential decision-making. Memory Gym \citep{pleines2025memory}, for example, introduces endless, procedurally generated environments to test agents’ memory effectiveness over unbounded horizons. MazeBench \citep{dao2025alphamaze} shifts the focus to LLMs, using tokenized maze representations, reinforcement learning, and chain-of-thought prompting to evaluate step-by-step spatial reasoning in small-size mazes. MazeEval \citep{einarsson2025mazeeval} isolates pure spatial reasoning by requiring LLMs to navigate mazes using only coordinate and distance-to-wall feedback, without visual input. Finally, MAPF has been proposed as a structured LLM benchmark, highlighting the unique difficulties of multi-agent coordination, long-horizon planning, and symbolic map understanding \citep{chen2025solving}.

Despite these advances, the results of existing maze benchmarks reveal that LLMs and RL agents struggle with the combinatorial demands of multi-agent coordination, especially in environments with a high degree in path deviations and obstacles \citep{chen2025solving, godin2025amaze}. Spatial and long-horizon reasoning present core challenges, with LLMs often failing to build robust internal representations and implement effective solutions towards global task completion. Further, structural limitations, such as context window size and lack of scalable memory-architectures hinder performance on large or complex mazes. Our work directly addresses these research gaps by introducing a unified, dynamic, and scalable framework that expands the limitations of prior LLM-based maze solving approaches, by supporting long-horizon task completion, active inference-based optimization loops, and real-time performance assessment for LLM-based agents.
\vspace{-2mm}
\paragraph{Reflective Instances in Multi-Agent Settings.} To orchestrate multi-agent interactions, previous research, such as \citep{shinn_reflexion_2023, chen_agentverse_2024, zhuge_language_2024}, has introduced reflective mechanisms that optimize agent-to-agent, and/or agent-to-environment coordination. Bo et al. \citep{bo_reflective_2025} introduce the COPPER framework, which implements a fine-tuned LLM to critique and refine outputs of primary agents, drawing on a reward mechanism to assess each agents overall contribution to over task success, and helping agents to perform on specific baseline benchmarks, including HotPotQA \citep{yang2018hotpotqa}, GSM8K \citep{cobbe2021training}, and `Checkmate in One Move' \citep{keskar2021checkmate}. Similarly, Nayak et al. \citep{nayak_long-horizon_2024} implement a plan-act-correct-verify mechanism to help an LLM-guided robot agent to autonomously navigate a 3D-environment; Leveraging visual feedback between environment and internal reflection instances to enhance navigational accuracy of the agent. Likewise, Xie et al. \citep{xie_teaching_2025} utilize reinforcement learning to improve an LLM-agent's ability to critique its own work, improving relative performance by 106\% on coding benchmarks. Whereas Ding et al. \citep{ding_multi-agent_2024} propose an asynchronous communication framework to optimize decision-capabilities in MAS, reducing error margins due to sequence-related circular dependencies in linear workflows. 
\vspace{-2mm}
\paragraph{Agents Partial Observability Limitations.}
While several authors have assessed the benefits of reflective instances to overcome agent's partial observability problem \citep{nayak_long-horizon_2024, madaan_self-refine_2023}, this strand of research is closely aligned with successes in agent-to-agent reinforcement learning pipelines \citep{shinn_reflexion_2023}. Expanding on these previous approaches, but aiming for emergent performance without reinforcement-driven validation, Ke et al.\citep{ke_mas-zero_2025} discuss an emergent zero-supervision MAS framework to overcome agent's partial observability problems and solve tasks at advanced complexity, by implementing a reflective meta-level instance that optimizes agent-to-agent sequences locally and need-based. In parallel, work such as \citep{huang_adasociety_2024} and \citep{li_adaptive_2025} introduces dynamic social graphs, or graph-attention paradigms \citep{niu_multi-agent_2021, chang_evince_2025} to support emergent multi-agent interaction in non-linear problem settings. While further, evolving orchestration approaches have been proposed \citep{zhuge_language_2024, chang_sagallm_2025, li_adaptive_2025, dang_multi-agent_2025} that formalize agent interaction as a directed graph, with a central orchestrator dynamically selecting and sequencing agent activations based on evolving task states, yielding more compact and efficient collaboration patterns. However, while these approaches are able to handle tasks at advanced complexity, they fall short in task completion that require a high number of steps across long-term planning and problem solving. To address these gaps, more recent work has suggested active inference principles, a neuroscience grounded paradigm that can be leveraged to aid agent's reasoning capabilities and drive task completion successes via quantified feedback principles \citep{yeganeh_deep_2025}. 
\vspace{-2mm}
\paragraph{Applications of Active Inference in MAS.}
We integrate active inference principles by  implementing a reflective benchmarking mechanisms to capture agents' \textit{free energy} \citep{suri_surprise_2022, walters_fe_risks_2025, yeganeh_deep_2025} and to control for system-wide entropy metrics and enhance long-term adaptability by nudging the system to reduce its error rates. There is a long tradition of benchmark-driven optimization in MAS research, which has found special traction in multi-agent reinforcement learning (MARL) contexts, such as \citep{niu_multi-agent_2021, ding_multi-agent_2024, assos_maximizing_2024, jiang_adaptive_2024}. Authors justify the need for dynamic benchmarks to optimize agent behavior in highly dynamic and hard to predict environments. For example, Suri et al. \citep{suri_surprise_2022} stabilize multi-agent interactions by collectively minimizing free energy across agent distributions, effectively reducing the occurrence of unexpected states. While, Ruiz-Serra et al.\citep{ruiz-serra_factorised_2025} integrate active inference frameworks to incorporate agents’ assumption about other agents’ internal states for strategic decision-making in iterative scenarios. In this work, we draw on recent advances on reflective mechanisms in LLM-based agent-optimization \citep{chang_evince_2025, chang_sagallm_2025, bo_reflective_2025, dang_multi-agent_2025} and merge the approach with active inference dynamic-optimization mechanisms, as demonstrated in \citep{prakki_active_2025, ruiz-serra_factorised_2025, yeganeh_deep_2025}. Our system uniquely synthesizes these reflection-based and information-theoretic insights by implementing a reflective, benchmark-driven orchestration instance to continuously evaluate and enhance agent-to-agent and agent-to-environment interactions.

In summary, Orchestrator unifies three key advances in the above literature: First, we implement a modular 'cell-structured' graph design, embedding planning, execution, and orchestration as explicit computational stages within each 'cell'. Second, we introduce a reflective benchmarking mechanism, using active inference principles to monitor FE-grounded intra- and inter-cell performance metrics to continuously assess and adapt local agent routines in light of global progress. Lastly, we leverage these performance metrics to foster dynamic adjustments of LLM's internal policy and prompt design, nudging agents to adjust their behavior if they encounter local solution minima or exhibit other behavior that is stalling progress. In this manner, and optimizing for both immediate and longitudinal performance, Orchestrator’s cell-based task-execution design empowers benchmark-driven coordination between components of an agent ensemble, proactively reducing error potential while improving task accuracy at the local and systemic level.

\section{Orchestrator Framework}
\label{sec:orchestrator}

\subsection{Graph-Based and Dynamic Multi-Agent Architecture}
We formalize the \textbf{Orchestrator} framework shown in \autoref{fig:orchestrator-framework} as a unified graph-based architecture that enables dynamic behavioral adaptation and real-time coordination.
The system's update sequence is modeled as a directed graph in \autoref{fig:orch-sub1}, while each system state at iteration \textit{t} is represented by the agentic 'cell' architecture, illustrating agent interactions and coordination dynamics as depicted in \autoref{fig:orch-sub2}.

Mathematically, the Orchestrator framework is defined as,
\begin{equation}
    \text{Orchestrator} = G(N, E, F)
\label{eq:graph}
\end{equation}
\noindent Where
$N = N_{\mathrm{plan}} \cup N_{\mathrm{exec}} \cup N_{\mathrm{orch}}$ denotes the complete node set, with each node corresponding to an agent. The framework consists of three node types: the planning node ($N_{\mathrm{plan}}$), the execution node ($N_{\mathrm{exec}}$), and the orchestration node ($N_{\mathrm{orch}}$). 
For this work, we implement a single orchestrator cell-instance, comprising one planning node and one orchestration node, with the remaining $(N-2)$ nodes instantiated as execution nodes. 
The plan node, $N_{\mathrm{plan}} = \{P\}$, provides a sequence of strategic action steps that guide execution nodes. 
Execution nodes, $N_{\mathrm{exec}} = \{e_1, e_2, \dots, e_n, \dots, e_{N-2}\}$, powered by LLMs, interpret the policy prompt and act in the environment. Where directed edges $E \subseteq (N \times N)$ are associated with routing functions, $G_r(t) = \{{g_r}_{(1;t)}, {g_r}_{(2;t)}, \dots, {g_r}_{(n;t)}, \dots, {g_r}_{(N-2;t)}\}$, that define define interaction pathways among agents at iteration $t$.
The action performance of each execution node is continuously evaluated by its variational free energy (VFE) following active inference principles (described in more detail in section \ref{sec:free_energy_benchmarking}). This ensures that actions are selected to maximize information gain, where high uncertainty corresponds to active exploration, while driving agents toward the global objective of maze completion. While execution nodes have knowledge of other agent's explored maze junctions at all times, the orchestration node, $N_{\mathrm{orch}} = \{O\}$, serves as a communication hub and global memory, allowing execution nodes to share additional information (such as other agent error rates or dead end detections) indirectly.

At the beginning of the maze exploration, the execution nodes $N_{\mathrm{exec}}$ and the orchestration node $N_{\mathrm{orch}}$ are initialized with distinct states, denoted as $S^e_t$ and $S^o_t$, respectively, which are dynamically updated over time. For the purpose of demonstration, and while future implementations may incorporate adaptive planning, we further initialize planning node $P$ with a preset sequence of \textit{k} steps, which is passed as loop of actionable instructions into execute nodes at $S_{n;(t-1)}$ (see section \ref{appendix:state_optimization_algorithm} in the Appendix).
Each execute node $E$ is equipped with its own, active-inference-based optimization function $f_{N}$, described in detail in the section \ref{sec:free_energy_benchmarking}.
Orchestration node $O$ maintains global state including the information of states of all execute nodes, while the state of each execute node is defined as a local state. 
At the start of each iteration $t$, each execute node considers (i) its own state and (ii) the states of other execute nodes updated by the orchestrator node and stored in its temporary state $S_t$ and then takes actions following its internal policy $\pi$ and the actionable-guidance $P_{plan}$ of length $k$. Each node $e^n_{S(t-1)}$ then loops through a predefined sequence of $k$ steps until plan completion or intervention triggers are met (Appendix \ref{appendix:state_optimization_algorithm}).

After completion of each step k, the weight of each execute node, written as $\Delta w$, is updated by calculating its VFE through the respective optimization function, $\{f_1, f_2, \dots, ... f_n, \dots, f_{N-2}\}$. The outcomes of these computations are then used to dynamically inject guidance instructions into temporary local states of execution nodes, $\{\hat{S}^e_{1;(t-1)}, \hat{S}^e_{2;(t-1)}, \dots, \hat{S}^e_{n;(t-1)}, \dots, \hat{S}^e_{(N-2);(t-1)}\}$.  Next, at the execution layer, the temporary local states are consolidated into updated local states,  $\{S^e_{1;t}, S^e_{2;t}, \dots, S^e_{n;t}, \dots, S^e_{(N-2);t}\}$. This system state is then passed to the orchestration node $O$, which reviews agent progress and provide direct optimization recommendations for execution agents. These recommendations are delivered through dynamic prompt injections that update the execution agents' internal policy, $\pi^E_{t-1}$, as the global state $S^O_t$ is passed to the next iteration. The detailed optimization computation for agents is explained in section \ref{sec:free_energy_benchmarking} and we provide a sequential overview of how the orchestration update sequence operates during maze exploration in \autoref{fig:orchestrator_extended_schematic}.

\begin{figure}[ht]
    \fbox{%
    \includegraphics[width=0.95\linewidth]{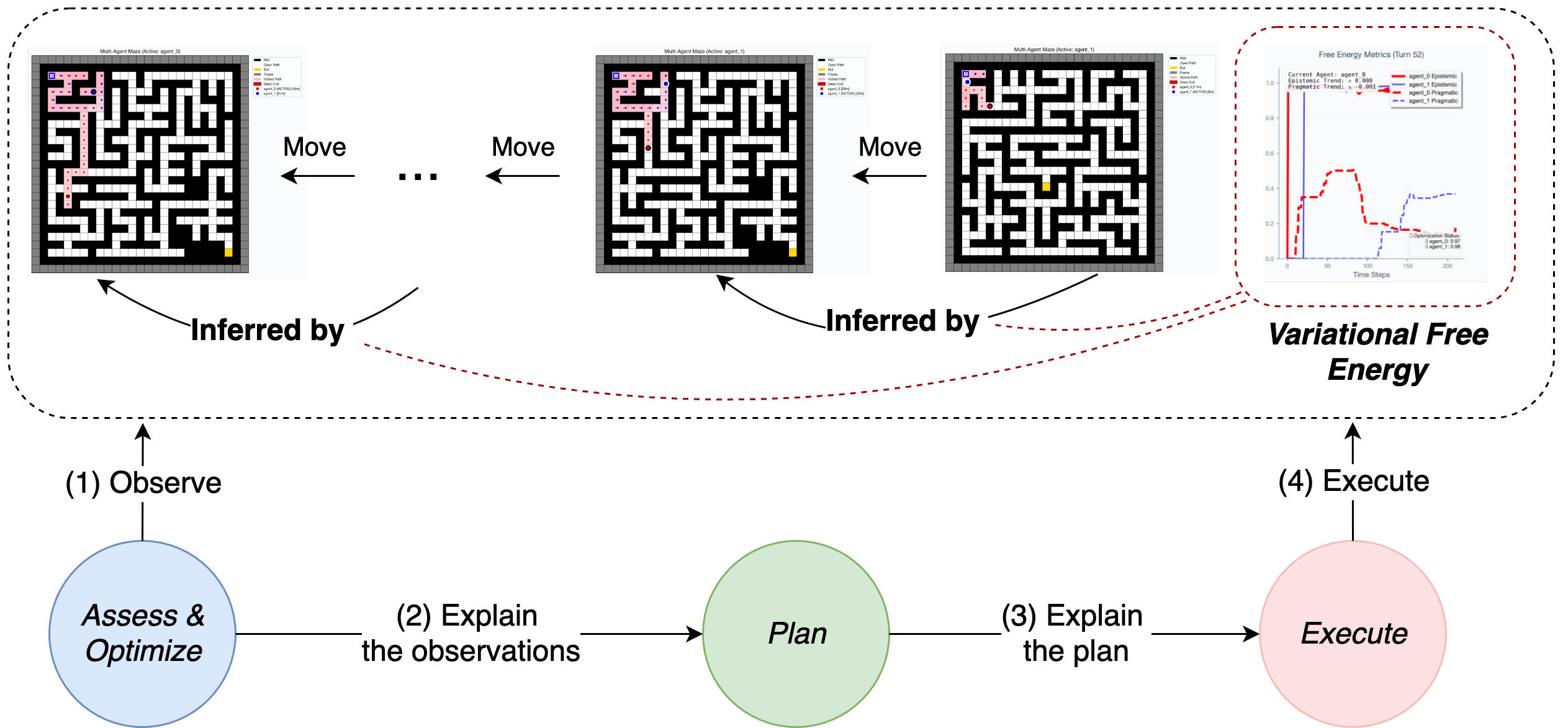}
    }
    \caption{Schematic representation of Orchestrator's decision-making cycle while solving a medium-difficulty maze-puzzle across n-steps.
    \vspace{-3mm}
    }
    \label{fig:orchestrator_extended_schematic}
\end{figure}
\subsection{Active Inference Benchmarking and Performance Assessment}\label{sec:free_energy_benchmarking}
Building on active inference principles, we reformulate the VFE objective. Instead of solely minimizing surprise to reduce deviations from the model predictions, we define the objective as a balanced trade-off: maximizing agents' realized information gain to encourage active learning, while offsetting this with an explicit cost function that captures coordination demands and behavioral efficiency. We provide the extended formalism to this approach in section \ref{appendix:active_inference} in the Appendix and operationalize it as follows:
\vspace{-1mm}
\paragraph{Epistemic Uncertainty.} We quantify epistemic uncertainty through measuring information entropy between consecutive states $\{S_t, S_{t+1}\}$, providing a real-time estimate of each agent's rate of actual information gain
\begin{equation}
    U_{\mathrm{epistemic}}(n,t,k) = - H[S_{n,t,k} \mid S_{n-1,t-1,k-1}]
\end{equation}
where $H[S_{n,t,k}]$ denotes the normalized Shannon entropy of the message output for agent $n$ at iteration $t$ and step $k$. 
The entropy is computed over individual message tokens $j$ as:
\begin{equation}
H_{\mathrm{tokens}}(n,t,k) = -\sum_{j \in K_{\mathrm{message}}} p_j(k) \log p_j(k)    
\end{equation}
\vspace{-5mm}
\paragraph{Accuracy Cost Assessment.} As a counterbalance to the agent's epistemic uncertainty, we operationalize the accuracy principle of VFE as cost term. While in principle the VFE accuray term corresponds to the expected negative log-likelihood of observed outcomes under the generative model, direct access to these likelihoods is unavailable due to the underlying LLM architectures. As described in section \ref{appendix:active_inference}, we therefore approximate accuracy cost using a behavioral proxy that captures behavioral efficiency and coordination at each step $k$:
\begin{equation}
C_{\mathrm{accuracy}}(n,t,k) = \sum_{j=1}^{5} w_j \cdot R_j(n,t,k)   
\end{equation}
where the static weights $w_j = 0.20$ for all $j \in \{1,2,3,4,5\}$ equally weight a set of five predefined risk components:
\begin{enumerate}[nosep] 
    \item \textbf{Movement Efficiency}: $R_{1}(n,t,k) = 1 - \frac{\text{total moves}}{\text{total move attempts}}$
    
    \item \textbf{Exploration Efficiency}: $R_{2}(n,t,k) = 1 - \frac{\text{unique positions visited}}{\text{total moves}}$
    
    \item \textbf{Backtracking Patterns}: $R_{3}(n,t,k) = \text{backtrack ratio} + 1.5 \cdot \text{oscillation penalty}$

    \item \textbf{Dead-End Recognition}: $R_4(n, t,k) = 1 - \frac{\text{dead-end revisits}}{\text{total moves}}$

    \item \textbf{Oscillation Avoidance}: $R_5(n,t,k) = 1 - \frac{\text{unique positions in recent moves}}{\text{recent move count}}$
\end{enumerate}
\vspace{-1mm}
\paragraph{Behavioral Optimization Function.} To normalize outputs, we cap both uncertainty and cost terms at $\pm 2.0$ and define the variational free energy $\mathcal{F}_n(t,k)$ for each agent $n$ at iteration $t$ and step $k$ as:
\begin{equation}\label{eq:free energy}
    F_{n}(t,k) = U_{\mathrm{epistemic}}(n,t,k) - C_{\mathrm{accuracy}}(n,t,k)
\end{equation}
The subtractive formulation reflects our optimization principle: high epistemic uncertainty signals productive information gain (positive contribution), whereas high pragmatic costs penalizes counterproductive behaviors including oscillation patterns, redundant exploration, or movement failures (negative contribution).
\vspace{-2mm}
\paragraph{Performance Policies.} Based on free energy outcomes, the system assigns agents to one of four performance categories, where threshold variables $\vartheta_1$ and $\vartheta_2$ have been deliberately assigned to match best performance using grid-search as determining method (see section \ref{appendix:threshold_grid_search} in the Appendix).

\begin{itemize}[nosep]
    \item \textbf{High Epistemic Drive, Low Accuracy Cost} ($U_{\mathrm{epistemic}} > 0.6$, $C_{\mathrm{accuracy}} < 0.4$): Effective exploration with efficient execution

    \item \textbf{High Epistemic Drive, High Accuracy Cost} ($U_{\mathrm{epistemic}} > 0.6$, $C_{\mathrm{accuracy}} > 0.4$): Active discovery but inefficient execution 

    \item \textbf{Low Epistemic Drive, Low Accuracy Cost} ($U_{\mathrm{epistemic}} < 0.6$, $C_{\mathrm{accuracy}} < 0.4$): Consistent execution but limited exploration

    \item \textbf{Low Epistemic Drive, High Accuracy Cost} ($U_{\mathrm{epistemic}} < 0.6$, $C_{\mathrm{accuracy}} > 0.4$): Poor exploration and inefficient execution
\end{itemize}
\vspace{-1mm}

\paragraph{Dynamic Weight Modulation.}
Each performance category triggers adjustments to a set of behavioral weights:
\begin{equation}
    \mathbf{w}_n(t,k)
    = \{\,w_{\mathrm{explore}}(t,k),\,
            w_{\mathrm{exploit}}(t,k),\,
            w_{\mathrm{coordinate}}(t,k),\,
            w_{\mathrm{backtrack}}(t,k)\}
\end{equation}
which are updated as:
\begin{equation}
    \mathbf{w}_n(t,k) = \mathbf{w}_{\mathrm{base}} + \Delta\mathbf{w}(F_n(t,k), \nabla F_n(t,k))
\end{equation}
where $\Delta\mathbf{w}$ accounts for both current free-energy and its temporal gradient, $\nabla\mathcal{F}_n(t,k)$, thus incorporating  predictive dynamics. 
For example, agents in the expressing high epistemic drive but high costs, are assigned increased $w_{\mathrm{exploit}}$ weights to improve execution efficiency, while agents with low drive but effective cost management, receive higher $w_{\mathrm{explore}}$ weights to encourage broader environmental exploration. Further details on movement reward scoring and dynamic weight updates are provided in sections \ref{appendix:state_optimization_algorithm} and \ref{appendix:movement_scores}. Prompt design of agents are presented in sections \ref{appendix:execution_agent_instructions} and \ref{appendix:orchestrator_agent_instructions}.
\vspace{-1mm}

\section{Experiments}
\label{experiments}

We evaluate our approach on a suite of synthetic maze environments designed to stress-test reasoning and coordination across long horizon task settings. We draw on the AMaze benchmark \citep{godin2025amaze}, which is designed to procedurally generate challenging maze environments and assess generalization ability of RL-agents across (long-horizon) task settings. The algorithmic design of our AMaze implementation is provided in Appendix section \ref{appendix:maze_generation_procedure}.
\vspace{-2mm}
\paragraph{Challenge.} We consider three maze difficulty levels (\textit{easy, medium}, \textit{hard}), each instantiated with five unique mazes. As defined by the AMaze benchmark, mazes differ in length, branching factor, and required coordination. A run is considered successful if one of $n$ execution agents reaches the maze exit within the maximum step budget. As maximum step budget we allocate a heuristic of two-and-a-half times the number of tiles per maze configuration, as well as a maximum duration of 7200 seconds, if one of these conditions is reached, timeout is initialized and the maze exploration is considered as failed. 
\vspace{-2mm}
\paragraph{Agents.} To ensure efficiency and responsiveness under real-world deployment constraints—where computational resources and budget are critical—we instantiate our agents using state-of-the-art, but compact, fast-inference LLMs (specifically GPT-4.1-nano and GPT-5-nano). For the purpose of demonstration, we present experiments with n=2 execution agents, only. While preliminary tests have showcased the feasibility of n=1 or n=3 agents, we consider n=2 a balanced trade-off in terms of efficiency, speed, and resource-allocation for the purpose of maze exploration. Further, while mixed-model approaches are possible, we constrain the orchestrator, when enabled, to being identical to the execution model, privileging a homogeneous approach.
\vspace{-2mm}
\paragraph{Baselines and Experiment Configurations.} We evaluate three core experimental configurations to systematically assess the impact of benchmarking and orchestration. As a floor baseline, we implement a random walk agent: a memory-enhanced, single-agent policy that self-selects valid moves at each step, with no access to FE-benchmarking, or orchestration support. Second, we introduce FE-benchmarking, providing agents with real-time feedback and dynamic weight adjustments, based on their performance. The third configuration adds an orchestrator node in addition to FE-benchmarking, to test for the models ability to facilitate higher-level coordination between agents.  To save compute, we omit the random walk for hard-level difficulty as chance of success is considerably low, as well as assessment of the FE + orchestration configuration on easy-level difficulty, as chances for success are considerably high.\vspace{-2mm}
\paragraph{Execution and Evaluation Metrics.} We execute at least 10 runs per configuration and level of difficulty, across 15 mazes in a balanced setting, yielding a minimum of 150 runs in total. For each configuration, we report key metrics such as success rate, total number of steps taken, number of failed moves, and total costs (normalized API token usage in dollars). We avoid wall-clock time measures, as provider rate limits and real-time changes on OpenAI API demand distort the results. Further, we predeclare a precision target of $\pm 15$ percentage points (pp) for success rate CIs. Runs are increased until this target is reached.

\section{Results and Discussion}\label{results_and_discussion}
\begin{table}[th]
\centering
\setlength{\tabcolsep}{6pt}
\vspace{-5mm}
\caption{Success rates with Wilson 95\% confidence intervals (CI) and corresponding half-widths in percentage points (pp) per model configurations and maze difficulty levels.}
\label{table_results_comparison}
\resizebox{\linewidth}{!}{%
\begin{tabular}{l l c c c c c c}
\toprule
Configuration & Difficulty & $\#$ of Runs & Successes & Success Rate (\%) & 95\% CI Lower & 95\% CI Upper & Half-width (pp) \\
\midrule
gpt-4.1-nano (Solo) & easy & 34 & 11 & 32.35 & 19.13 & 49.16 & 15.01 \\
gpt-4.1-nano (Solo) & medium & 33 & 10 & 30.3 & 17.38 & 47.34 & 14.98 \\
gpt-4.1-nano + FE Benchmark only & easy & 10 & 10 & 100.0 & 72.25 & 100.0 & 13.88 \\
gpt-4.1-nano + FE Benchmark only & medium & 36 & 26 & 72.22 & 56.01 & 84.15 & 14.07 \\
gpt-4.1-nano + FE Benchmark only & hard & 26 & 22 & 84.62 & 66.47 & 93.85 & 13.69 \\
gpt-4.1-nano + FE + Orchestration Node  & medium & 25 & 25 & 100.0 & 86.68 & 100.0 & 6.66 \\
gpt-4.1-nano + FE + Orchestration Node  & hard & 32 & 23 & 71.88 & 54.63 & 84.44 & 14.9 \\
gpt-5-nano (Solo) & easy & 10 & 0 & 0.0 & 0.0 & 27.75 & 13.88 \\
gpt-5-nano (Solo) & medium & 11 & 0 & 0.0 & 0.0 & 25.88 & 12.94 \\
gpt-5-nano + FE Benchmark only & easy & 10 & 10 & 100.0 & 72.25 & 100.0 & 13.88 \\
gpt-5-nano + FE Benchmark only & medium & 25 & 20 & 80.0 & 60.87 & 91.14 & 15.14 \\
gpt-5-nano + FE Benchmark only & hard & 36 & 23 & 63.89 & 47.58 & 77.52 & 14.97 \\
gpt-5-nano + FE + Orchestration Node & medium & 24 & 20 & 83.33 & 64.15 & 93.32 & 14.59 \\
gpt-5-nano + FE + Orchestration Node & hard & 30 & 23 & 76.67 & 59.07 & 88.21 & 14.57 \\
\bottomrule
\end{tabular}
}
\end{table}
We present our results in \autoref{table_results_comparison}. For solo agent ensembles without benchmark and orchestration nodes, we find poor model performance in runs on easy-level (32,35\% success rate for GPT-4.1-nano; 0\% success rate for GPT-5-nano) and medium-level difficulty (30,03\% for GPT-4.1-nano;  0\% for GPT-5-nano). In contrast, Orchestrator strongly outperforms solo agent ensembles on easy (100\% for both GPT-nano models) and medium-difficulty levels (100\% for GPT-4.1.-nano and 83.33\% for for GPT-5-nano), accounting for a threefold increase in success rates for GPT-4.1-nano configurations and an increase by a factor of 3,33 and 2,77 respectively for easy and medium levels in GPT-5-nano configurations. Further, we find that just incorporating FE-benchmarks, already substantially improves model performance (medium: 72.22\% for GPT-4.1-nano, 80.0\% for GPT-5-nano; hard: 84.62\% for GPT-4.1-nano, 63.89\% for GPT-5-nano). Lastly, adding orchestration improves performance for 3 out of 4 configurations  (medium: 100\% for GPT-4.1-nano, 83.33\% for GPT-5-nano; hard: 71.88\% for GPT-4.1-nano, 76.67\% for GPT-5-nano). Surprisingly, GPT-4.1-nano with FE benchmarks outperforms the same model with added orchestration on hard mazes (84.62\% vs. 71.88\%). One possible explanation is that orchestration, while generally beneficial, can introduce additional reasoning overhead in high-complexity environments, exceeding the model’s effective planning horizon \citep{shojaee_illusion_2025}. In such cases, more streamlined reasoning—guided by FE benchmarks alone—may yield better results \citep{yeganeh_deep_2025, belcak2025smalllanguagemodelsfuture}. This suggests that while additional orchestration instances improve task-completion accuracy for most scenarios \citep{dang_multi-agent_2025, zhang_optimizing_2025, chang_sagallm_2025}, their actual utility may depend on correctly applying the ensemble composition of agents to match the specific task at hand. Additional results concerning model cost-effectiveness, completion efficiency, and convergence intervals per configuration are listed in sections \ref{appendix:performance_review}, and \ref{appendix:ci_convergence} in the Appendix.
\begin{figure}[t!]{
\vspace{-5mm}
\centering
\includegraphics[width=0.95\linewidth]{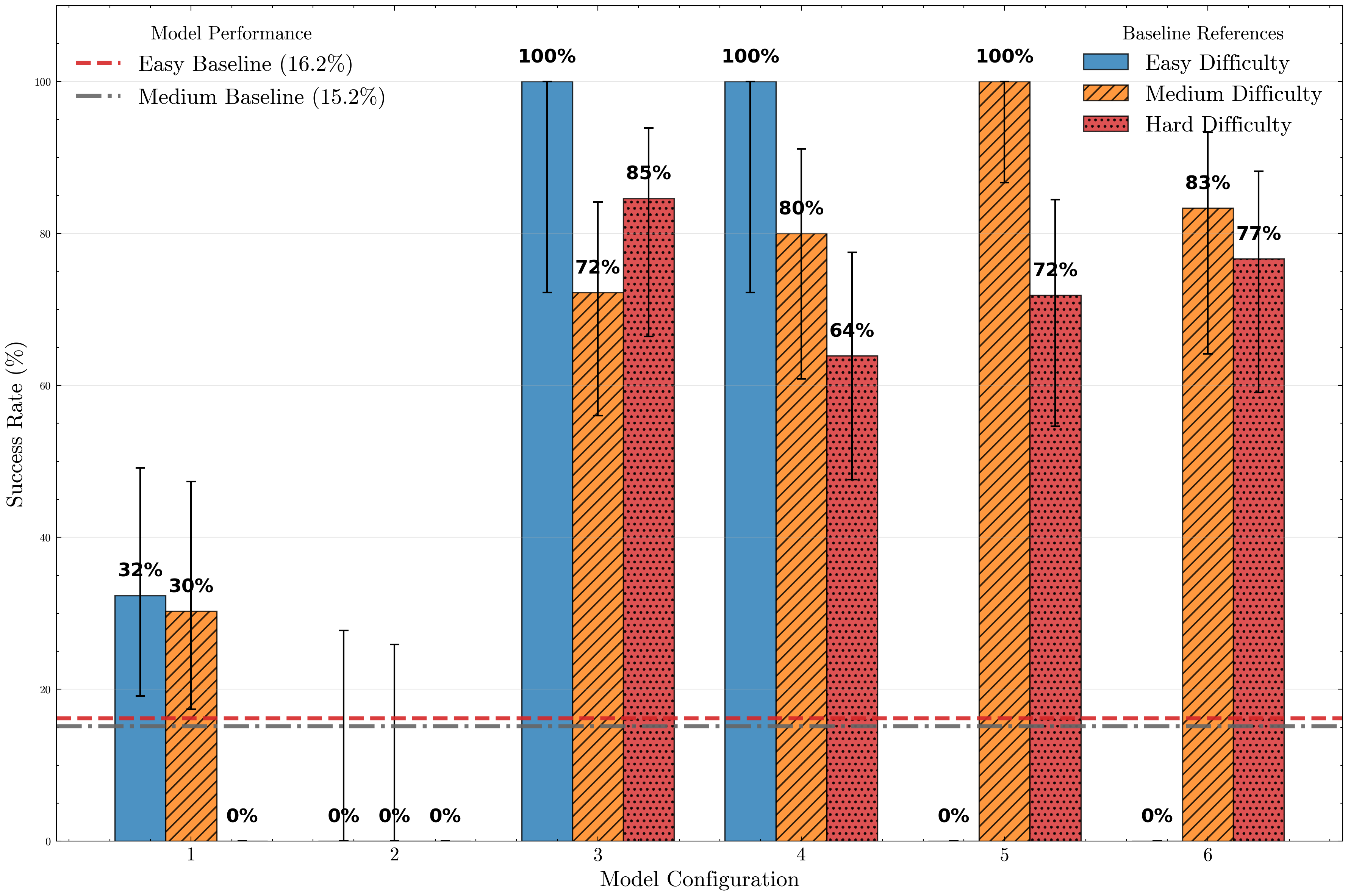}
\label{fig:success_rates}
}
\vspace{0mm}
\caption{Success rate and Wilson CI ranges by model configuration and difficulty as shown in \autoref{table_results_comparison}. Numerical identifiers for configurations are as follows: 1) GPT-4.1-nano (Solo); 2) GPT-5-nano (Solo); 3)  GPT-4.1-nano + FE Benchmark only; 4) GPT-5-nano + FE Benchmark only; 5) GPT-4.1-nano + FE + Orchestration Node; 6) GPT-5-nano + FE + Orchestration Node.}
\vspace{-5mm}
\label{fig:model_constellation_performance}
\end{figure}
\vspace{0mm}
\section{Conclusion}
\label{conclusion}
\vspace{0mm}
This paper introduced Orchestrator, a unified active inference-based framework for multi-agent coordination in long-horizon environments. Motivated by the limitations of existing long-horizon, maze solving approaches—which often fail to capture the challenges of memory-retention, adaptability, and long-horizon coordination—Orchestrator integrates dynamic feedback, reflective benchmarking, and modular orchestration into a single, scalable architecture. 

Our results highlight several key contributions: First, we propose a novel active-inference-based architecture for multi-agent coordination that supports real-time adaptation, memory-driven collaboration, and scalable reasoning. Second, we show that this approach achieves strong performance even with lightweight, fast-inference models suitable for real-world deployment, but is able to accommodate more sophisticated reasoning agents in alignment with rising task complexity. Third, we demonstrate that the combination of active-inference benchmarking and orchestration substantially improves both the reliability and efficiency of agent teams, particularly in complex, long-horizon tasks. By bridging the gap between static benchmarks and adaptive, feedback-driven orchestration, our work contributes to the debate on foundational frameworks for robust, autonomous, and LLM-based multi-agent systems capable of addressing long-horizon challenges in real-world production settings. 

Nevertheless several limitations remain: As the present analysis is restricted to synthetic maze environments and small agent teams, the ability of Orchestrator to address long-horizon tasks in more heterogeneous settings and across other problem domains remains to be explored. Further, to assess essential deployment factors in terms of scalability, cost, and performance, framework performance should be extensively tested for addressing higher-complexity tasks while using larger LLM models. Lastly, given the fact that our architecture achieves high level performance with smaller-size, high inference-speed LLMs, we see great potential for deploying Orchestrator using open-source models. The feasibility of these approaches should be tested in future iterations.

\newpage

\bibliographystyle{unsrt}
\bibliography{refs}

\begin{thebibliography}{10}

\bibitem{chen_agentverse_2024}
Weize Chen, Yusheng Su, Jingwei Zuo, Cheng Yang, Chenfei Yuan, Chi-Min Chan, Heyang Yu, Yaxi Lu, Yi-Hsin Hung, Chen Qian, Yujia Qin, Xin Cong, Ruobing Xie, Zhiyuan Liu, Maosong Sun, and Jie Zhou.
\newblock {AgentVerse}: {Facilitating Multi-Agent Collaboration and Exploring Emergent Behaviors in Agents}.
\newblock 2024.

\bibitem{li_more_2024}
Junyou Li, Qin Zhang, Yangbin Yu, Qiang Fu, and Deheng Ye.
\newblock More {Agents} {Is} {All} {You} {Need}, October 2024.
\newblock arXiv:2402.05120 [cs].

\bibitem{zhang_exploring_2024}
Jintian Zhang, Xin Xu, Ningyu Zhang, Ruibo Liu, Bryan Hooi, and Shumin Deng.
\newblock Exploring {Collaboration} {Mechanisms} for {LLM} {Agents}: {A} {Social} {Psychology} {View}, May 2024.
\newblock arXiv:2310.02124 [cs].

\bibitem{liu_lessons_2025}
Yuanzhe Liu, Ryan Deng, Tim Kaler, Xuhao Chen, Charles~E. Leiserson, Yao Ma, and Jie Chen.
\newblock Lessons {Learned}: {A} {Multi}-{Agent} {Framework} for {Code} {LLMs} to {Learn} and {Improve}, 2025.
\newblock Version Number: 1.

\bibitem{he_llm-based_2025}
Junda He, Christoph Treude, and David Lo.
\newblock {LLM}-{Based} {Multi}-{Agent} {Systems} for {Software} {Engineering}: {Literature} {Review}, {Vision} and the {Road} {Ahead}, July 2025.
\newblock arXiv:2404.04834 [cs].

\bibitem{Xu_vis-analysis_2025}
Chao Xu, Qi~Zhang, Baiyan Li, Anmin Wang, and Jingsong Bao.
\newblock Visual analysis of time series data for multi-agent systems driven by large language models.
\newblock In {\em Proceedings of the 3rd International Conference on Signal Processing, Computer Networks and Communications}, SPCNC '24, page 427–431, New York, NY, USA, 2025. Association for Computing Machinery.

\bibitem{xu_multi-agent_2024}
Liming Xu, Sara Almahri, Stephen Mak, and Alexandra Brintrup.
\newblock Multi-{Agent} {Systems} and {Foundation} {Models} {Enable} {Autonomous} {Supply} {Chains}: {Opportunities} and {Challenges}.
\newblock {\em IFAC-PapersOnLine}, 58(19):795--800, 2024.
\newblock Publisher: Elsevier BV.

\bibitem{xu_implementing_2024}
Liming Xu, Stephen Mak, Maria Minaricova, and Alexandra Brintrup.
\newblock On {Implementing} {Autonomous} {Supply} {Chains}: a {Multi}-{Agent} {System} {Approach}, June 2024.
\newblock arXiv:2310.09435 [cs].

\bibitem{alon_multiagent_2020}
Tal Alon, Magdalen Dobson, Ariel Procaccia, Inbal Talgam-Cohen, and Jamie Tucker-Foltz.
\newblock Multiagent {Evaluation} {Mechanisms}.
\newblock {\em Proceedings of the AAAI Conference on Artificial Intelligence}, 34(02):1774--1781, April 2020.

\bibitem{chen_optima_2025}
Weize Chen, Jiarui Yuan, Chen Qian, Cheng Yang, Zhiyuan Liu, and Maosong Sun.
\newblock Optima: {Optimizing} {Effectiveness} and {Efficiency} for {LLM}-{Based} {Multi}-{Agent} {System}, February 2025.
\newblock arXiv:2410.08115 [cs].

\bibitem{yao_hdflow_2024}
Wenlin Yao, Haitao Mi, and Dong Yu.
\newblock {HDFlow}: {Enhancing} {LLM} {Complex} {Problem}-{Solving} with {Hybrid} {Thinking} and {Dynamic} {Workflows}, September 2024.
\newblock arXiv:2409.17433 [cs].

\bibitem{prakki_active_2025}
Rithvik Prakki.
\newblock Active {Inference} for {Self}-{Organizing} {Multi}-{LLM} {Systems}: {A} {Bayesian} {Thermodynamic} {Approach} to {Adaptation}, January 2025.
\newblock arXiv:2412.10425 [cs].

\bibitem{iqbal_actor-attention-critic_2018}
Shariq Iqbal and Fei Sha.
\newblock Actor-{Attention}-{Critic} for {Multi}-{Agent} {Reinforcement} {Learning}.
\newblock 2018.

\bibitem{liu_grounded_2024}
Zeyang Liu, Xinrui Yang, and Shiguang Sun.
\newblock Grounded {Answers} for {Multi}-agent {Decision}-making {Problem} through {Generative} {World} {Model}.
\newblock 2024.

\bibitem{ding_multi-agent_2024}
Ziluo Ding, Zeyuan Liu, Zhirui Fang, Kefan Su, Liwen Zhu, and Zongqing Lu.
\newblock Multi-{Agent} {Coordination} via {Multi}-{Level} {Communication}.
\newblock 2024.

\bibitem{erdogan_plan-and-act_2025}
Lutfi~Eren Erdogan, Nicholas Lee, Sehoon Kim, Suhong Moon, Hiroki Furuta, Gopala Anumanchipalli, Kurt Keutzer, and Amir Gholami.
\newblock Plan-and-{Act}: {Improving} {Planning} of {Agents} for {Long}-{Horizon} {Tasks}, April 2025.
\newblock arXiv:2503.09572 [cs].

\bibitem{zhuge_language_2024}
Mingchen Zhuge, Wenyi Wang, Louis Kirsch, Francesco Faccio, Dmitrii Khizbullin, and Jürgen Schmidhuber.
\newblock Language {Agents} as {Optimizable} {Graphs}, August 2024.
\newblock arXiv:2402.16823 [cs].

\bibitem{nayak_long-horizon_2024}
Siddharth Nayak, Adelmo~Morrison Orozco, Jackson Zhang, Darren Chen, Aditya Kapoor, Eric Robinson, Karthik Gopalakrishnan, James Harrison, Brian Ichter, Anuj Mahajan, and Hamsa Balakrishnan.
\newblock Long-{Horizon} {Planning} for {Multi}-{Agent} {Robots} in {Partially} {Observable} {Environments}.
\newblock 2024.

\bibitem{chang_sagallm_2025}
Edward~Y. Chang and Longling Geng.
\newblock {SagaLLM}: {Context} {Management}, {Validation}, and {Transaction} {Guarantees} for {Multi}-{Agent} {LLM} {Planning}, July 2025.
\newblock arXiv:2503.11951 [cs].

\bibitem{li_adaptive_2025}
Boyi Li, Zhonghan Zhao, Der-Horng Lee, and Gaoang Wang.
\newblock Adaptive {Graph} {Pruning} for {Multi}-{Agent} {Communication}, June 2025.
\newblock arXiv:2506.02951 [cs].

\bibitem{rein2024gpqa}
David Rein, Betty~Li Hou, Asa~Cooper Stickland, Jackson Petty, Richard~Yuanzhe Pang, Julien Dirani, Julian Michael, and Samuel~R Bowman.
\newblock {GPQA: A Graduate-Level Google-Proof Q\&A Benchmark}.
\newblock In {\em First Conference on Language Modeling}, 2024.

\bibitem{kapoor_ai_2025}
Sayash Kapoor, Benedikt Stroebl, Zachary~S. Siegel, Nitya Nadgir, and Arvind Narayanan.
\newblock {AI} {Agents} {That} {Matter}.
\newblock {\em Transactions on Machine Learning Research}, June 2025.

\bibitem{dang_multi-agent_2025}
Yufan Dang, Chen Qian, Xueheng Luo, Jingru Fan, Zihao Xie, Ruijie Shi, Weize Chen, Cheng Yang, Xiaoyin Che, Ye~Tian, Xuantang Xiong, Lei Han, Zhiyuan Liu, and Maosong Sun.
\newblock Multi-{Agent} {Collaboration} via {Evolving} {Orchestration}, May 2025.
\newblock arXiv:2505.19591 [cs].

\bibitem{zhang_optimizing_2025}
Enhao Zhang, Erkang Zhu, Gagan Bansal, Adam Fourney, Hussein Mozannar, and Jack Gerrits.
\newblock Optimizing {Sequential} {Multi}-{Step} {Tasks} with {Parallel} {LLM} {Agents}, July 2025.
\newblock arXiv:2507.08944 [cs].

\bibitem{xiao_tradingagents_2025}
Yijia Xiao, Edward Sun, Di~Luo, and Wei Wang.
\newblock {TradingAgents}: {Multi}-{Agents} {LLM} {Financial} {Trading} {Framework}, June 2025.
\newblock arXiv:2412.20138 [q-fin].

\bibitem{walters_fe_risks_2025}
Michael Walters, Rafael Kaufmann, Justice Sefas, and Thomas Kopinski.
\newblock Free energy risk metrics for systemically safe ai: Gatekeeping multi-agent study, 2025.
\newblock arXiv:2502.04249 [cs.AI].

\bibitem{shojaee_illusion_2025}
Parshin Shojaee, Iman Mirzadeh, Keivan Alizadeh, Maxwell Horton, Samy Bengio, and Mehrdad Farajtabar.
\newblock The {Illusion} of {Thinking}: {Understanding} the {Strengths} and {Limitations} of {Reasoning} {Models} via the {Lens} of {Problem} {Complexity}, July 2025.
\newblock arXiv:2506.06941 [cs].

\bibitem{dao2025alphamaze}
Alan Dao and Dinh~Bach Vu.
\newblock {AlphaMaze: Enhancing Large Language Models' Spatial Intelligence via GRPO}.
\newblock {\em arXiv preprint arXiv:2502.14669}, 2025.

\bibitem{parr_active_inference_2022}
Thomas Parr, Giovanni Pezzulo, and Karl~J. Friston.
\newblock {\em Active Inference: The Free Energy Principle in Mind, Brain, and Behavior}.
\newblock The MIT Press, 03 2022.

\bibitem{bo_reflective_2025}
Xiaohe Bo, Zeyu Zhang, Quanyu Dai, Xueyang Feng, Lei Wang, Rui Li, Xu~Chen, and Ji-Rong Wen.
\newblock Reflective {Multi}-{Agent} {Collaboration} based on {Large} {Language} {Models}.
\newblock 2025.

\bibitem{ruiz-serra_factorised_2025}
Jaime Ruiz-Serra, Patrick Sweeney, and Michael~S. Harré.
\newblock Factorised {Active} {Inference} for {Strategic} {Multi}-{Agent} {Interactions}, May 2025.
\newblock arXiv:2411.07362 [cs].

\bibitem{suri_surprise_2022}
Karush Suri, Xiao~Qi Shi, Konstantinos Plataniotis, and Yuri Lawryshyn.
\newblock Surprise {Minimizing} {Multi}-{Agent} {Learning} with {Energy}-based {Models}.
\newblock 2022.

\bibitem{yeganeh_deep_2025}
Yavar~Taheri Yeganeh, Mohsen Jafari, and Andrea Matta.
\newblock Deep {Active} {Inference} {Agents} for {Delayed} and {Long}-{Horizon} {Environments}, May 2025.
\newblock arXiv:2505.19867 [cs].

\bibitem{omidshafiei_deep_2017}
Shayegan Omidshafiei, Jason Pazis, Christopher Amato, Jonathan~P. How, and John Vian.
\newblock Deep {Decentralized} {Multi}-task {Multi}-{Agent} {Reinforcement} {Learning} under {Partial} {Observability}, July 2017.
\newblock arXiv:1703.06182 [cs].

\bibitem{linardakis_distributed_2024}
Manousos Linardakis, Iraklis Varlamis, and Georgios~Th. Papadopoulos.
\newblock Distributed {Maze} {Exploration} {Using} {Multiple} {Agents} and {Optimal} {Goal} {Assignment}.
\newblock {\em IEEE Access}, 12:101407--101418, 2024.

\bibitem{godin2025amaze}
Kevin Godin-Dubois, Karine Miras, and Anna~V Kononova.
\newblock {AMaze: An Intuitive Benchmark Generator for Fast Prototyping of Generalizable Agents}.
\newblock {\em Frontiers in Artificial Intelligence}, 8:1511712, 2025.

\bibitem{stern2019multi}
Roni Stern, Nathan Sturtevant, Ariel Felner, Sven Koenig, Hang Ma, Thayne Walker, Jiaoyang Li, Dor Atzmon, Liron Cohen, TK~Kumar, et~al.
\newblock {Multi-Agent Pathfinding: Definitions, Variants, and Benchmarks}.
\newblock In {\em Proceedings of the International Symposium on Combinatorial Search}, volume~10, pages 151--158, 2019.

\bibitem{foead_systematic_2021}
Daniel Foead, Alifio Ghifari, Marchel~Budi Kusuma, Novita Hanafiah, and Eric Gunawan.
\newblock A {Systematic} {Literature} {Review} of {A}* {Pathfinding}.
\newblock {\em Procedia Computer Science}, 179:507--514, 2021.

\bibitem{tjiharjadi_systematic_2022}
Semuil Tjiharjadi, Sazalinsyah Razali, and Hamzah~Asyrani Sulaiman.
\newblock A {Systematic} {Literature} {Review} of {Multi}-agent {Pathfinding} for {Maze} {Research}.
\newblock {\em Journal of Advances in Information Technology}, 13(4), 2022.

\bibitem{liu_cooperative_2025}
Ning Liu, Sen Shen, Xiangrui Kong, Hongtao Zhang, and Thomas Bräunl.
\newblock Cooperative {Hybrid} {Multi}-{Agent} {Pathfinding} {Based} on {Shared} {Exploration} {Maps}, March 2025.
\newblock arXiv:2503.22162 [cs].

\bibitem{pleines2025memory}
Marco Pleines, Matthias Pallasch, Frank Zimmer, and Mike Preuss.
\newblock {Memory Gym: Towards Endless Tasks to Benchmark Memory Capabilities of Agents}.
\newblock {\em Journal of Machine Learning Research}, 26(6):1--40, 2025.

\bibitem{einarsson2025mazeeval}
Hafsteinn Einarsson.
\newblock {MazeEval: A Benchmark for Testing Sequential Decision-Making in Language Models}.
\newblock {\em arXiv preprint arXiv:2507.20395}, 2025.

\bibitem{chen2025solving}
Weizhe Chen, Sven Koenig, and Bistra Dilkina.
\newblock Solving multi-agent path finding as an {LLM} benchmark: How, how good and why.
\newblock {\em Transactions on Machine Learning Research}, 2025.

\bibitem{shinn_reflexion_2023}
Noah Shinn, Federico Cassano, Edward Berman, Ashwin Gopinath, Karthik Narasimhan, and Shunyu Yao.
\newblock Reflexion: {Language} {Agents} with {Verbal} {Reinforcement} {Learning}, October 2023.
\newblock arXiv:2303.11366 [cs].

\bibitem{yang2018hotpotqa}
Zhilin Yang, Peng Qi, Saizheng Zhang, Yoshua Bengio, William~W Cohen, Ruslan Salakhutdinov, and Christopher~D Manning.
\newblock {HotpotQA: A Dataset for Diverse, Explainable Multi-Hop Question Answering}.
\newblock {\em arXiv preprint arXiv:1809.09600}, 2018.

\bibitem{cobbe2021training}
Karl Cobbe, Vineet Kosaraju, Mohammad Bavarian, Mark Chen, Heewoo Jun, Lukasz Kaiser, Matthias Plappert, Jerry Tworek, Jacob Hilton, Reiichiro Nakano, et~al.
\newblock {Training Verifiers to Solve Math Word Problems}.
\newblock {\em arXiv preprint arXiv:2110.14168}, 2021.

\bibitem{keskar2021checkmate}
Nitish~Shirish Keskar.
\newblock {Checkmate in One Move}.
\newblock \url{https://github.com/google/BIG-bench/blob/main/bigbench/benchmark_tasks/checkmate_in_one/README.md}, 2021.
\newblock Accessed: 2025-08-21.

\bibitem{xie_teaching_2025}
Zhihui Xie, Jie Chen, Liyu Chen, Weichao Mao, Jingjing Xu, and Lingpeng Kong.
\newblock Teaching {Language} {Models} to {Critique} via {Reinforcement} {Learning}, February 2025.
\newblock arXiv:2502.03492 [cs].

\bibitem{madaan_self-refine_2023}
Aman Madaan, Niket Tandon, Prakhar Gupta, Skyler Hallinan, Luyu Gao, Sarah Wiegreffe, Uri Alon, Nouha Dziri, Shrimai Prabhumoye, Yiming Yang, Shashank Gupta, Bodhisattwa~Prasad Majumder, Katherine Hermann, Sean Welleck, Amir Yazdanbakhsh, and Peter Clark.
\newblock Self-{Refine}: {Iterative} {Refinement} with {Self}-{Feedback}, May 2023.
\newblock arXiv:2303.17651 [cs].

\bibitem{ke_mas-zero_2025}
Zixuan Ke, Austin Xu, Yifei Ming, Xuan-Phi Nguyen, Caiming Xiong, and Shafiq Joty.
\newblock {MAS}-{ZERO}: {Designing} {Multi}-{Agent} {Systems} with {Zero} {Supervision}, May 2025.
\newblock arXiv:2505.14996 [cs].

\bibitem{huang_adasociety_2024}
Yizhe Huang, Xingbo Wang, Hao Liu, Fanqi Kong, Aoyang Qin, Min Tang, Song-Chun Zhu, Mingjie Bi, Siyuan Qi, and Xue Feng.
\newblock {AdaSociety}: {An} {Adaptive} {Environment} with {Social} {Structures} for {Multi}-{Agent} {Decision}-{Making}.
\newblock 2024.

\bibitem{niu_multi-agent_2021}
Yaru Niu, Rohan Paleja, and Matthew Gombolay.
\newblock Multi-{Agent} {Graph}-{Attention} {Communication} and {Teaming}.
\newblock 2021.

\bibitem{chang_evince_2025}
Edward~Y. Chang.
\newblock {EVINCE}: {Optimizing} {Multi}-{LLM} {Dialogues} {Using} {Conditional} {Statistics} and {Information} {Theory}, January 2025.
\newblock arXiv:2408.14575 [cs].

\bibitem{assos_maximizing_2024}
Angelos Assos, Yuval Dagan, and Constantinos Daskalakis.
\newblock Maximizing utility in multi-agent environments by anticipating the behavior of other learners, July 2024.
\newblock arXiv:2407.04889 [cs].

\bibitem{jiang_adaptive_2024}
Ruichen Jiang, Ali Kavis, Qiujiang Jin, Sujay Sanghavi, and Aryan Mokhtari.
\newblock Adaptive and {Optimal} {Second}-order {Optimistic} {Methods} for {Minimax} {Optimization}.
\newblock 2024.

\bibitem{belcak2025smalllanguagemodelsfuture}
Peter Belcak, Greg Heinrich, Shizhe Diao, Yonggan Fu, Xin Dong, Saurav Muralidharan, Yingyan~Celine Lin, and Pavlo Molchanov.
\newblock Small language models are the future of agentic ai, 2025.

\bibitem{Jaynes_1957}
E.~T. Jaynes.
\newblock Information theory and statistical mechanics.
\newblock {\em Phys. Rev.}, 106:620--630, May 1957.

\end{thebibliography}

\newpage 
\appendix
\section{Technical Appendices and Supplementary Material}
\subsection{Orchestrator Update Algorithm} \label{appendix:state_optimization_algorithm}
\begin{algorithm}[H]
\caption{Multi-Agent Active Inference Maze Solver}
\begin{algorithmic}[1]
\Require Maze $\mathcal{M}$, start $s_0$, target $\tau$, number of execute agents $(N-2)$
\Require Static plan per iteration $P = [p_1,p_2,\dots,p_k,\dots,p_K]$ (length $K$ steps)
\State \textbf{Initialize:} 

    \State $S^O_0 \gets \text{OrchestratorNode.initialize}(\mathcal{M}, s_o, \tau)$ 
    \Comment{Orchestrator state $S^O_0$ contains all execute nodes' states (e.g. initial positions).}
    \State $target\_found \gets False$, $t \gets 1$

\Statex 

\While{\textbf{not} $target\_found$}  \Comment{Iterate until target is found}
    \State $\pi_{t-1} \gets \text{OrchestratorNode.encodePolicy}(S^O_{t-1}, P)$ 
    \Comment{Policy prompt combining plan and global state $S^O_{t-1}$.}
    \For{\textbf{each} execute node $n = 1$ \textbf{to} $N-2$ \textbf{and not} $target\_found$}
        \State $\hat{S}^e_{n;(t-1)} \gets f_n(S^O_{t-1}, \pi_{t-1}, S^e_{n;(t-1)})$ 
        \Comment{Execute node $n$ integrates policy prompt into its own state.}
        \For{\textbf{each} step $k = 1$ \textbf{to} $K$ \textbf{in plan}}
            \If{step $k$ is "LookAround"} 
                \State Node $n$ observes surroundings and updates its local map/beliefs.
            \ElsIf{step $k$ is "SelectDirection"} 
                \State Node $n$ chooses the best direction to move (based on its observations).
            \ElsIf{step $k$ is "MarkDeadEnd"} 
                \State Node $n$ marks current position as dead-end in its memory (if applicable).
            \EndIf
            \If{step $k$ requires an actual move} 
                \State Node $n$ executes move in the decided direction. 
                \Comment{Directional action in the maze environment.}
                \State Node $n$ observes new state (e.g. new position and sensory inputs).
            \EndIf
            \State Compute variational free energy $F_n(t, k)$ for node $n$ at this step.
            \State $\Delta F_n(t, k) \gets F_n(t, k) - F_n(t-1, k)$ 
            \Comment{Change in variational free energy from previous step.}
            \State $\Delta w_n(t,k) \gets f_{\Delta}\!\big(F_n(t, k), \Delta F_n(t, k)\big)$ 
            \Comment{Gradient-like update for node $n$'s parameters considering $F_n(t, k)$ and $\Delta F_n(t, k)$.}
            \State $w_{base} \gets w_n(t,k-1) + \Delta w_n(t,k)$ 
            \Comment{Update node $n$'s internal parameters.}
            \State Update node $n$'s state $S^e_{n;t}$ (e.g. new position, updated memory).
            \If{Node $n$ has reached target}
                \State $target\_found \gets True$
                \State \textbf{break from both for-loops} 
                \Comment{Exit if the target is found.}
            \EndIf
        \EndFor
        \State $S^O_t \gets \text{OrchestratorNode.update}(S^e_{1;t}, S^e_{2;t}, \dots, S^e_{n;t}, \dots, S^e_{N-2;t})$ 
        \Comment{Orchestrator node updates its state with all execute nodes' new state.}
    \EndFor
    \State $t \gets t - 1$ \Comment{Increment time/iteration counter.}
\EndWhile
\State \textbf{Output:} Path or solution found by agents reaching the goal.
\end{algorithmic}
\end{algorithm}

\clearpage
\subsection{Orchestrator Cell Architecture Overview}\label{appendix:orchestrator-framework_large}

\begin{figure}[h]
    \centering
    \includegraphics[width=\linewidth]{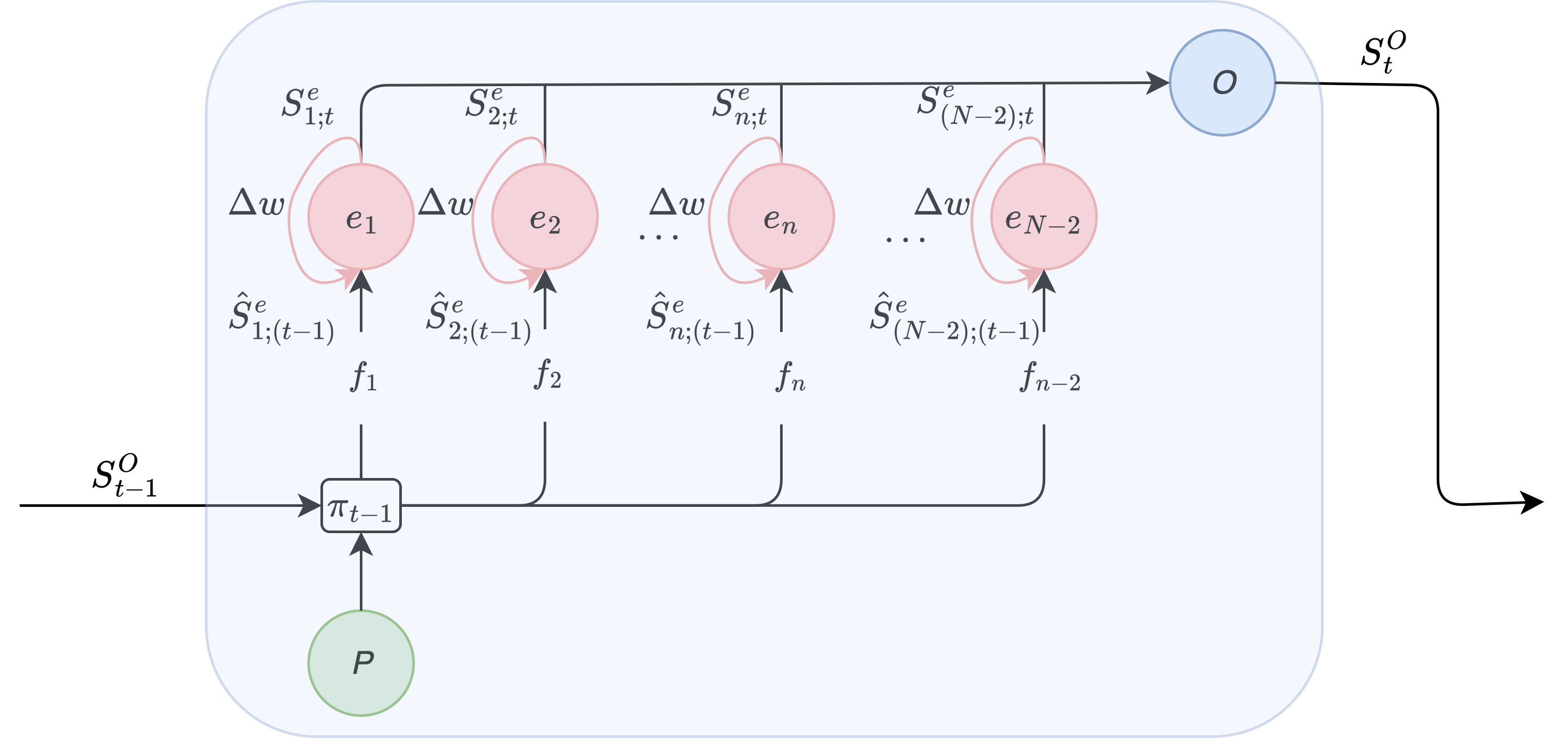}
    \caption{Orchestrator Cell Design - large-size reprint of \autoref{fig:orch-sub2}.}
\end{figure}

\subsection{Active Inference Calculation}\label{appendix:active_inference}
Drawing on active inference principles, we reformulate the variational free energy (VFE) objective. 
Rather than minimizing surprise to reduce deviation from a model's predictions, we cast the objective as a balanced process: maximizing expected Bayesian surprise to encourage active learning, while offsetting this with explicit penalty terms for coordination and navigation efficiency. 

Active inference, rooted in computational neuroscience, provides a principled framework for reasoning under uncertainty and has been shown to enhance agents' task performance through quantified feedback mechanisms \citep{yeganeh_deep_2025, parr_active_inference_2022}.
Given the assumption that agents operate under partial observability of the problem environment and risk becoming trapped in local minima during task optimization, we operationalize VFE to nudge agents towards active exploration and learning-driven behavior.

Following \citep{parr_active_inference_2022}, VFE is defined as,

\begin{align*}
VFE = F[Q,y] &= - \underbrace{\mathbb{E}_{Q(x)}[\ln P(y,x)]}_{\text{Energy}}
          - \underbrace{H[Q(x)]}_{\text{Entropy}} \\[6pt]
&= \underbrace{D_{\mathrm{KL}}[Q(x) \,||\, P(x)]}_{\text{Complexity}}
   - \underbrace{\mathbb{E}_{Q(x)}[\ln P(y|x)]}_{\text{Accuracy}} \\[6pt]
\end{align*}
where
\begin{itemize}
    \item $\mathbb{E}_{Q(x)}[\cdot]$ denotes the expectation under the approximate posterior $Q(x)$,
    \item $H[Q(x)]$ is the entropy of $Q(x)$,
    \item $D_{\mathrm{KL}}[Q(x)\,||\,P(x)]$ is the Kullback–Leibler (KL) divergence between $Q(x)$ and $P(x)$,
    \item $P(y|x)$ is the likelihood of $y$ given latent state $x$,
\end{itemize}

Expanding the entropy definition, we obtain:
\begin{align*}
F[Q,y] &= - \underbrace{\mathbb{E}_{Q(x)}[\ln P(y,x)]}_{\text{Energy}}
          - \underbrace{H[Q(x)]}_{\text{Entropy}} \\[6pt]
          &=  -\mathbb{E}_{Q(x)}[\ln P(y,x)] - \big(-\mathbb{E}_{Q(x)}[\ln Q(x)]\big) \\[6pt]
           &= -\mathbb{E}_{Q(x)}[\ln P(y|x) + \ln P(x)] + \mathbb{E}_{Q(x)}[\ln Q(x)] \\[6pt]
           &= \mathbb{E}_{Q(x)}\left[ \ln \frac{Q(x)}{P(x)} \right] - \mathbb{E}_{Q(x)}[\ln P(y|x)] \\[6pt]
           &= \underbrace{D_{\mathrm{KL}}[Q(x) \,||\, P(x)]}_{\text{Complexity}}
   - \underbrace{\mathbb{E}_{Q(x)}[\ln P(y|x)]}_{\text{Accuracy}} \\[6pt]
   \end{align*}

In practice, however, direct access to LLM internal prior and exact likelihoods is unavailable. To address this, we adopt Jaynes’s maximum entropy principle \citep{Jaynes_1957}, approximating the unknown posterior P(x) as uniform. Substituting into the KL term yields:

\begin{align*}
D_{\mathrm{KL}}[Q(x),||,P(x)]
&= \mathbb{E}_{Q(x)}[\ln Q(x)] + \ln(N),
\end{align*}
\noindent where $N$ denotes the support size of the uniform prior. Therefore, VFE simplifies to: 
\begin{align*}
F[Q,y] &= \underbrace{-H[Q(x)] + \ln(N)}_{\text{Actual Information Gain}}
- \underbrace{\mathbb{E}_{Q(x)}[\ln P(y|x)]}_{\text{Accuracy Term}}.
\end{align*}

We interpret the first component as a proxy for epistemic value (actual information gain), while the second acts a penalty on inaccurate predictions. 
Approximating the likelihood-based cost, we define:
\begin{align*}
   \text{Accuracy Term} =  - \mathbb{E}_{Q(x)}[\ln P(y|x)] \approx\; - [\ln P(y|x)] \\[6pt]
\end{align*}
Finally, casting this into our operational form for Orchestrator:
\[
\begin{aligned}
\text{VFE} \;\approx\; &- H[S_t \mid S_{t-1}] \;-\; \mathbb{E}_{Q(x)}[\ln P(S_{t-1}|x)] \\[6pt]
&= - H[S_t \mid S_{t-1}] \;+\; \mathbb{E}\big[\text{Accuracy Cost}(S_{t-1})\big] \\[6pt]
&=  U_{\mathrm{epistemic}} - C_{\mathrm{accuracy}}
\end{aligned}
\]
\noindent where $ U_{\mathrm{epistemic}}= - H[S_t \mid S_{t-1}]$ measures active information gain across states, and $C_{\mathrm{accuracy}}$ penalizes inaccurate predictions based on expected negative log-likelihood.  

\subsection{Movement Score Policy Updates}\label{appendix:movement_scores}
The dynamic weights $\mathbf{w}_n(t,k)$ derived from free energy assessment are 
operationalized through directional movement scoring functions that convert 
performance metrics into actionable spatial insights, which are passed to the agent as part of the dynamic policy updates at {$S^O_t$. For each execution node $e_n$ at iteration $t$ and step $k$, the system computes movement scores $M_n(d,t,k)$ for each feasible direction $d \in \{\text{north}, \text{south}, \text{east}, \text{west}\}$ as:}
{\begin{equation}
    M_n(d,t,k) = \sum_{i} w_i(t, k) \cdot \phi_i(d, S^e_{n;t})
\end{equation}}
where $\phi_i(d,S^e_{n;t})$ encodes exploration, efficiency, coordination, and 
backtracking factors for direction $d$ given the current local state $S^e_{n;t}$. 
These movement scores provide execution nodes with quantified directional 
preferences that integrate both individual performance optimization and 
system-wide coordination objectives, enabling the translation of abstract free 
energy metrics into concrete spatial decisions within the maze environment, towards exploring the maze with greater efficiency.

\newpage
\subsection{Determination of Threshold Variables for Agent Performance Policy Assessment}\label{appendix:threshold_grid_search}

\begin{figure}[h]
    \centering
    \includegraphics[width=0.85\linewidth]{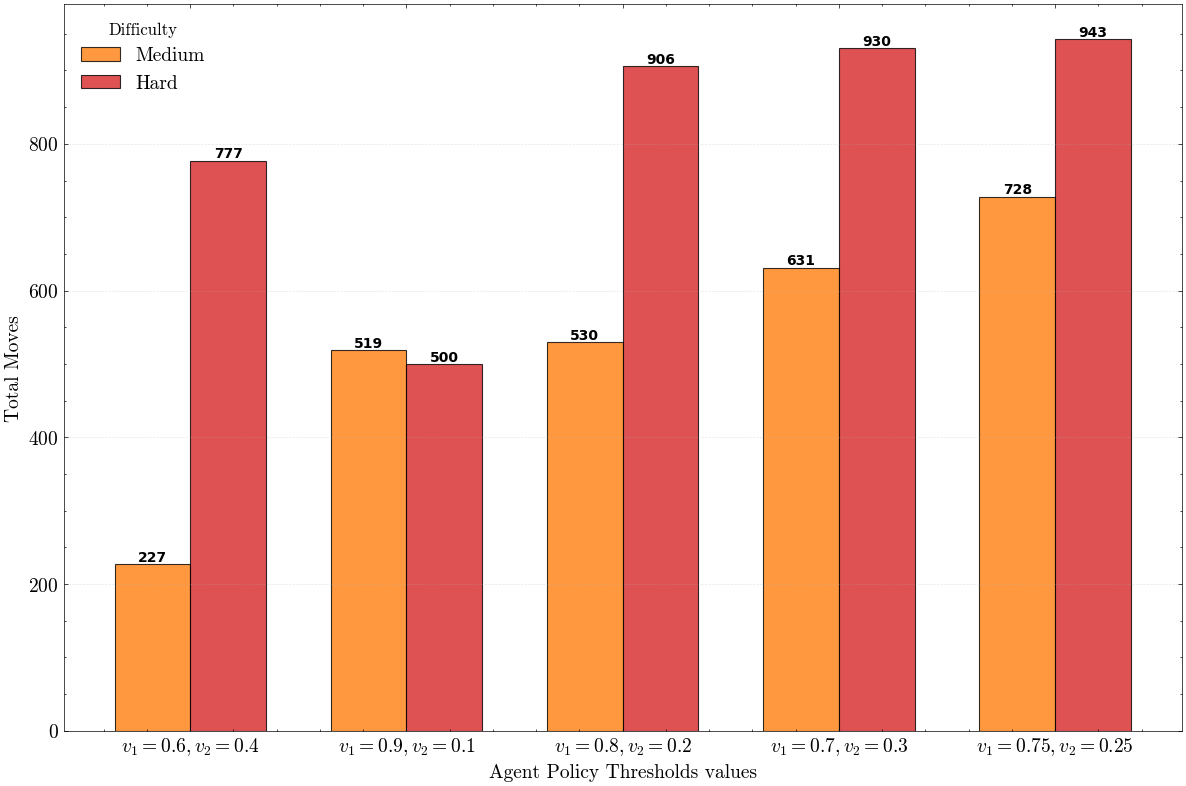}
    \caption{Results of grid-search to determine best threshold parameters for maximum performance of the Orchestrator framework.}
\end{figure}
 
 To identify optimal threshold values for agent performance policy assessment as discussed in section \ref{sec:free_energy_benchmarking}, we conducted a brief grid search over a set of threshold parameters $\vartheta_1$ (epistemic drive) and $\vartheta_2$ (accuracy cost) and assess performance in terms of total number of steps required to solve the maze, given the respective parameter setup across both difficulties (medium and hard). Maze-solving experiments were performed at two difficulty levels, with $n=3$ runs per setting. The results indicate that for medium-difficulty mazes, best performance is achieved with $\vartheta_1 = 0.6$ and $\vartheta_2 = 0.4$, while for hard mazes, optimal performance is observed at a lower accuracy cost threshold ($\vartheta_1 = 0.9, \vartheta_2 = 0.01$). However, total step count is slightly lower for the former setting ($\vartheta_1 = 0.6, \vartheta_2 = 0.04$) indicating subtly elevated performance to the latter. For consistency and comparability across all experiments in this paper, we adopt the higher-performance setting of $\vartheta_1 = 0.6$ and $\vartheta_2 = 0.4$ throughout. Future iterations should test the framework at additional threshold parameters for $\vartheta_1$ and $\vartheta_2$.

\subsection{Agent Tool Interface Specification} \label{appendix:tool_interface}

The execution agents operate within the maze environment through a structured tool interface that provides both environmental interaction capabilities and internal state management functions. This tool-based architecture ensures consistent action execution across all agents while maintaining proper state synchronization within the multi-agent framework.

\subsubsection{Spatial Navigation Tools}

The core navigation functionality is implemented through four directional movement tools: 
\texttt{move\_north()}, \texttt{move\_south()}, \texttt{move\_east()}, and \texttt{move\_west()}. 
Each tool attempts to execute a single-step movement in the specified cardinal direction and returns deterministic success/failure feedback. 
The tools operate on the agent's individual \texttt{MazeWrapper} instance, ensuring proper collision detection with walls and maze boundaries. 
Upon successful movement, the tool reports the agent's new position coordinates using matrix notation (row, column), while failed attempts provide specific failure reasons (e.g., blocked by wall, boundary violation).

\subsubsection{Environmental Perception and State Management}

The \texttt{get\_current\_view()} tool provides agents with local environmental perception through a structured observation that includes the agent's current position, available movement directions at a +1 tile horizon, exit proximity status, and a spatial representation of the immediate surroundings. 
This tool serves as the primary sensory input mechanism, enabling agents to make informed decisions based on their local environment state.

The \texttt{mark\_dead\_end()} tool allows agents to maintain persistent spatial memory by marking their current position as a dead end when specific confidence criteria are met. 
This tool supports the system's exploration efficiency by preventing redundant exploration of previously identified dead-end locations.

\subsubsection{Backtracking and Recovery Mechanisms}

The \texttt{start\_backtracking()} tool implements an automated recovery mechanism for agents that become stuck or require strategic repositioning. 
When invoked, this tool calculates the shortest path to the nearest unexplored opening using breadth-first search through the agent's movement history. 
The tool establishes a ``lock mode'' state that provides deterministic step-by-step navigation instructions until the target position is reached, ensuring reliable recovery from suboptimal positions. 
The backtracking mechanism operates exclusively through previously visited positions, maintaining consistency with the agent's explored knowledge while preventing navigation through unknown or potentially blocked areas. Upon completion of the backtracking sequence, agents automatically resume normal exploration behavior. This tool interface design ensures that while agents receive algorithmic assistance for basic spatial operations and state management, the high-level decision-making regarding which tools to use, when to initiate backtracking, and how to coordinate with other agents remains within the domain of the language model's reasoning capabilities.
\newpage
\subsubsection{Supplementary Performance Charts}\label{appendix:performance_review}
\begin{figure*}[ht]
    \centering
        \begin{minipage}{\linewidth}
        \centering
        \includegraphics[width=0.85\linewidth]{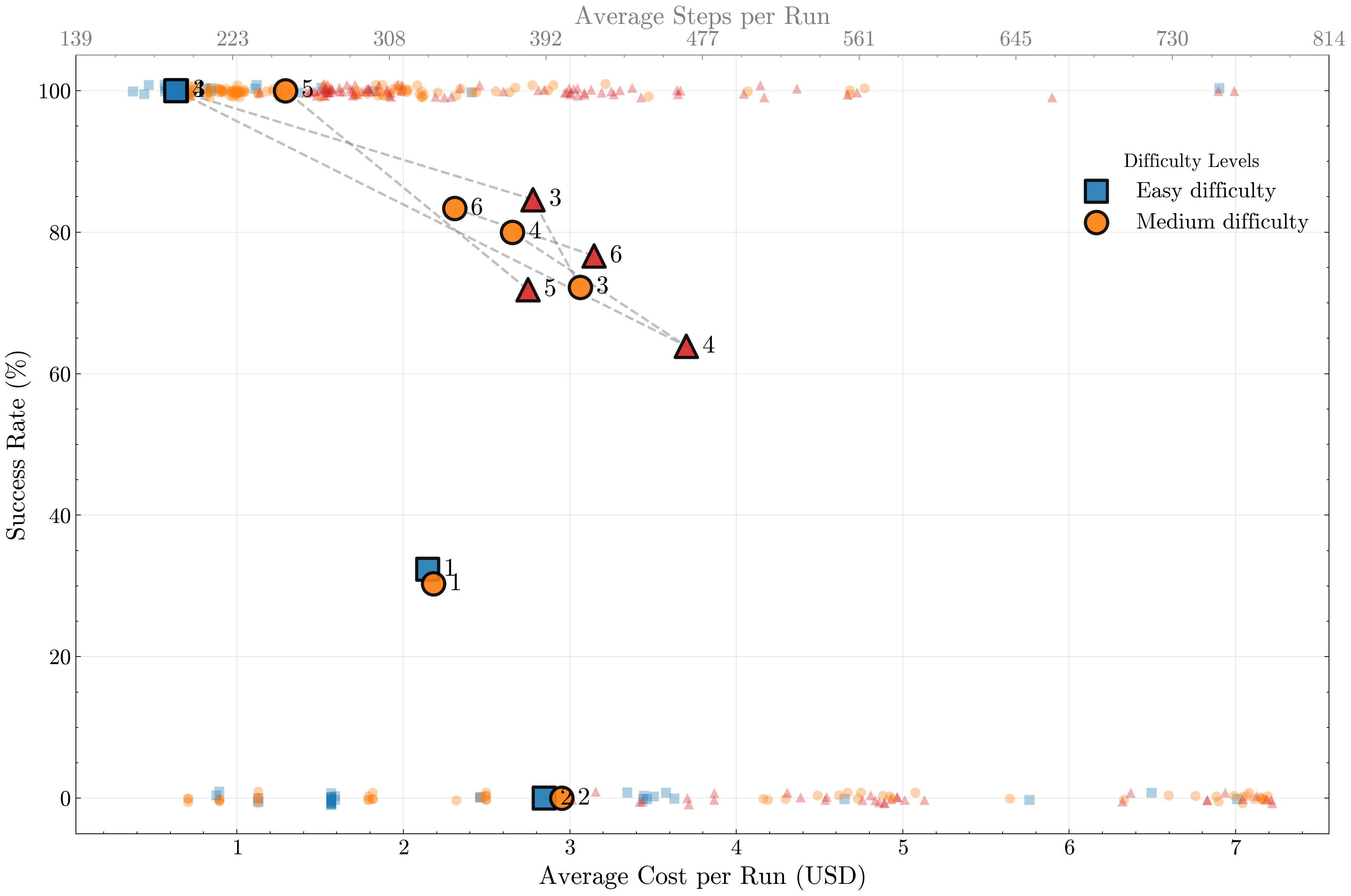}
        \caption{Cost-effectiveness of different configurations, showing the tradeoff between average run cost and success rate. Numerical identifiers for configurations are as follows: 1) GPT-4.1-nano (Solo); 2) GPT-5-nano (Solo); 3)  GPT-4.1-nano + FE Benchmark only 4) GPT-5-nano + FE Benchmark only; 5) GPT-4.1-nano + FE + Orchestration Node; 6) GPT-5-nano + FE + Orchestration Node}
        \label{fig:cost_analysis}
    \end{minipage}
    
    \vspace{1em} 
    
    \begin{minipage}{\linewidth}
        \centering
        \includegraphics[width=0.85\linewidth]{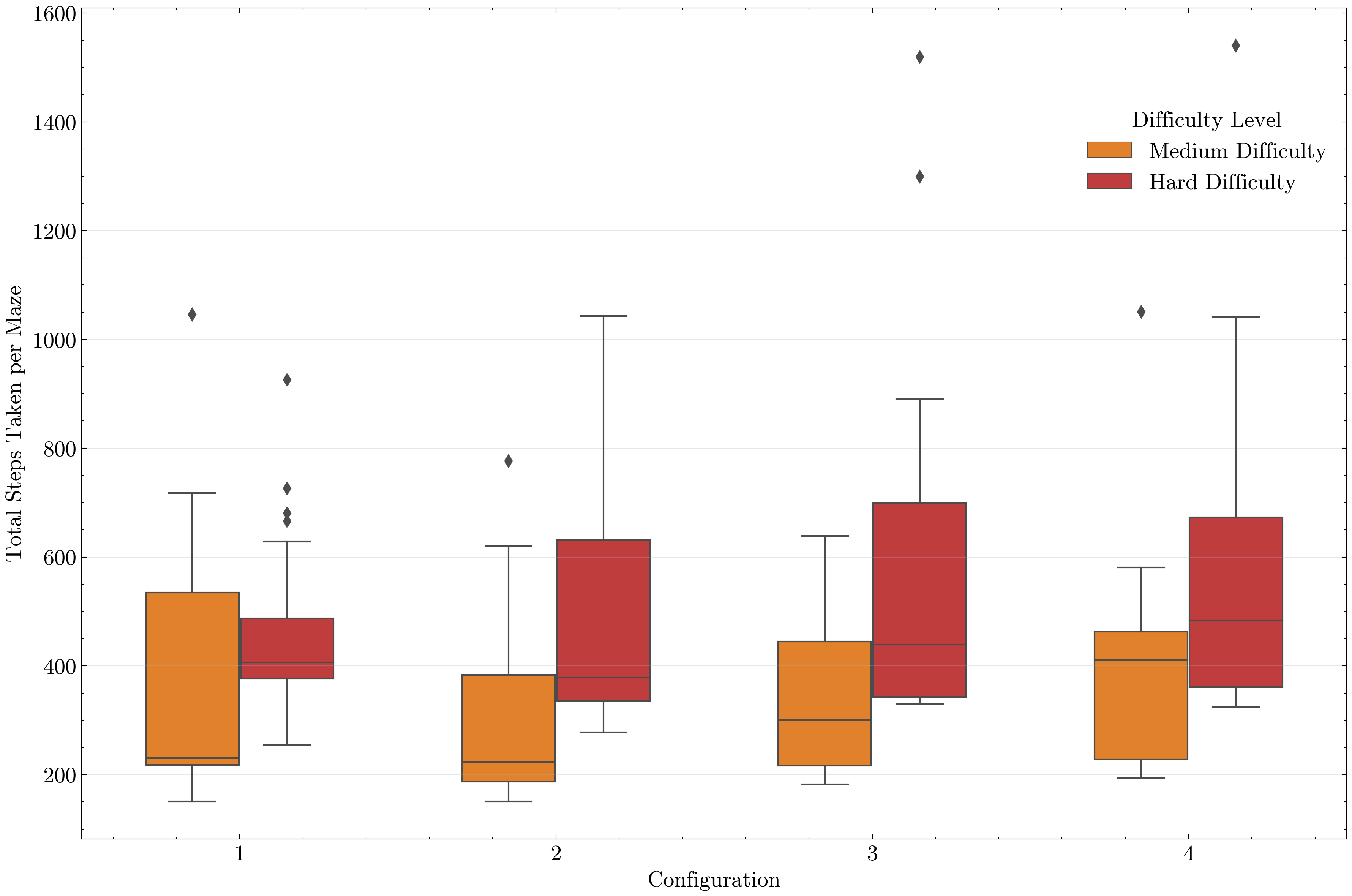}
        \caption{Distribution of steps taken to solve mazes, grouped by configuration and difficulty for medium- and hard-difficulty mazes using the orchestrator framework (successful runs only). Numerical identifiers for configurations are as follows: 1)  GPT-4.1-nano + FE Benchmark only; 2) GPT-4.1-nano + FE + Orchestration Node; 3) GPT-5-nano + FE Benchmark only; 4) GPT-5-nano + FE + Orchestration Node. Easy level has been omitted due to negligibly small scale.}
        \label{fig:steps_distribution}
    \end{minipage}
\end{figure*}

\clearpage
\subsubsection{Confidence Interval Convergence across Model Configurations}\label{appendix:ci_convergence}
\begin{figure}[ht]
\centering
\fbox{%
\includegraphics[width=0.95\linewidth]{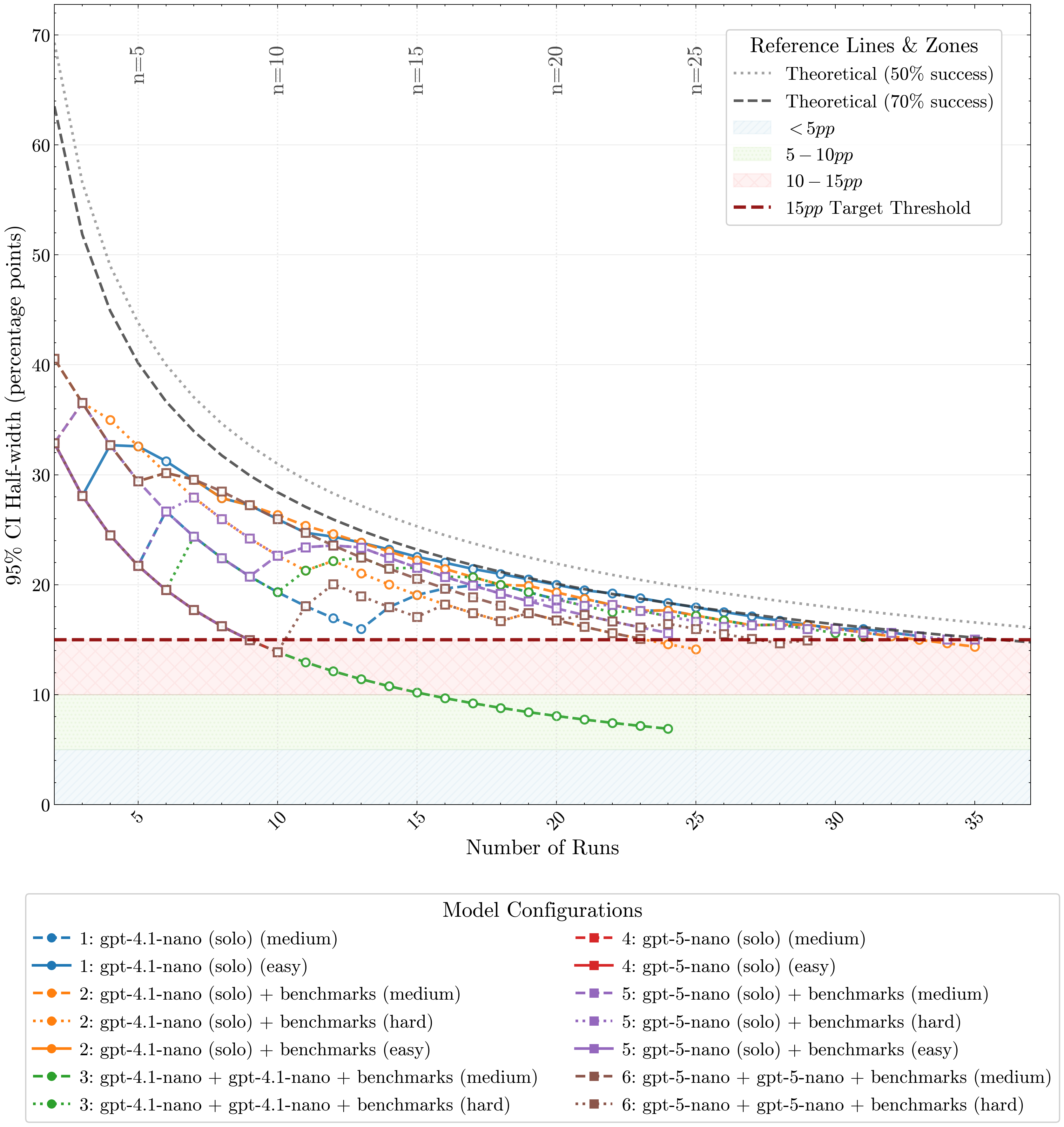}
}
\caption{Statistical convergence analysis demonstrating the stabilization of performance estimates with increasing sample size across different model configurations. Each line represents the 95\% confidence interval half-width for success rate estimates as a function of cumulative experimental runs, with different colors indicating distinct model constellations (execution model, orchestration model combinations). The y-axis shows the confidence interval half-width in percentage points, providing a direct measure of estimate precision. The shaded regions indicate target precision zones: green zone ($\leq 5$ percentage points) represents high precision suitable for reliable performance comparisons, yellow zone (5--15 percentage points) indicates moderate precision adequate for preliminary analysis, and white zone ($> 15$ percentage points). All configurations demonstrate asymptotic convergence behavior, with most achieving stable estimates ($\leq 10$ percentage points half-width) after 20--25 experimental runs. The reference lines show theoretical convergence bounds for different baseline success rates (50\%, 70\%) under Wilson score interval calculations, validating the identified convergence patterns.}
\label{fig:orchestrator_extended_schematic}
\end{figure}

\newpage
\subsection{Maze Execution Agent Prompt Template}

\paragraph{A. Model System Prompt}\label{appendix:execution_agent_instructions}
\begin{lstlisting}[style=prompt]
You are Agent {agent_id} in a collaborative maze escape.

CRITICAL RULE
- Must call exactly ONE tool per step.
- Only exception: mark_dead_end() is optional.

COORDINATE SYSTEM
- Matrix coordinates (row, col), not Cartesian.
- (3,5) means "row 3, column 5".
- NORTH = -row, SOUTH = +row, EAST = +col, WEST = -col.
- Maze is displayed like a spreadsheet grid, not a graph.

WEIGHTED DECISION SYSTEM
Guided by dynamic performance weights.

DECISION HIERARCHY (check in order)
1. Backtracking Lock Mode (override):
   If "BACKTRACKING LOCK MODE ACTIVE", immediately execute the required move.
   Ignore all other rules until cleared.
2. Coordinate with Teammates:
   Avoid teammate-explored areas unless no alternatives.
   Apply weight × {teammate_avoidance}.
3. Orchestrator & Optimization Guidance:
   Apply orchestrator corrections and optimization hints.
   Current weights: exploration={exploration_weight}, efficiency={efficiency_weight}.
4. Standard Backtracking Mode:
   If "BACKTRACKING ACTIVE", execute required move and skip other checks.
5. Oscillation Detection:
   If stuck looping (same 2–3 positions), call start_backtracking().
6. Safety Check:
   Never move into walls. If blocked everywhere, call start_backtracking().
7. Exploration Priority:
   Use weighted movement scores. Avoid dead ends unless necessary for backtracking.

AVAILABLE ACTIONS
- get_current_view() → Observe 3x3 surroundings
- move_north/south/east/west() → Advance one step
- mark_dead_end() → Optional, no args
- start_backtracking() → Return to nearest unexplored opening

DEAD END MARKING (threshold={dead_end_confidence})
Mark a cell as dead end only if:
(a) Only one possible move, leading back to visited tiles
(b) No unexplored directions remain
(c) Not currently backtracking
Skip marking if multiple unexplored paths exist, confidence < threshold,
or backtracking mode is active.

TURN STRUCTURE
1. get_current_view()
2. move_[direction]()
3. Optionally: mark_dead_end()

VICTORY CONDITION
- If "Maze Exit" found → return FINISH immediately.

FORBIDDEN
- Multiple tool calls per step
- Moving in loops
- Explaining reasoning
- Calling start_backtracking() when already backtracking
\end{lstlisting}

\paragraph{B. Execution-Context Message (runtime)}
\begin{lstlisting}[style=prompt]
EXECUTION CONTEXT – STEP {step_index + 1}

CURRENT STEP
- {current_step}

CURRENT STATE
- Position: {current_position}
- Available moves: {possible_moves}
- Current unexplored directions: {agent_unexplored_directions}
- Known unexplored openings: {known_openings}
- {_format_dynamic_modifiers(dynamic_prompts)}

WEIGHTED MOVEMENT ANALYSIS
- {movement_guidance}
- All direction scores: {score_details}
- Backtrack threshold: {weights.get('backtrack_threshold', 0.7):.2f}
- Dead end confidence: {dead_end_confidence:.2f}
  (threshold={weights.get('dead_end_confidence', 0.8):.2f})

BACKTRACKING STATUS
- Currently backtracking: {"YES" if is_backtracking else "NO"}
- Lock mode active: {"YES" if lock_mode else "NO"}
- WARNING: Do NOT call start_backtracking() if already backtracking

EXPLORATION STATUS
- Previously visited: {previously_visited_tiles}
- Dead ends marked: {len(marked_dead_ends) if marked_dead_ends else 0}
- Avoid backtracking to recent path unless other rules apply: {recent_positions}

OSCILLATION CHECK
- Recent movement pattern: {recent_positions}
- WARNING: If current position appears >2 times in recent pattern,
  call start_backtracking()

MULTI-AGENT COORDINATION
- AVOID returning to your previous position: 
  {previous_position if previous_position else "None"}
- Teammate recent positions (last 10, avoid if alternatives exist):
  {recent_other_positions[-10:] if len(recent_other_positions) >= 10 else recent_other_positions}
- Teammate explored junctions/dead ends (avoid if alternatives exist):
  {strategic_waypoints}

GUIDANCE
- Orchestrator: {agent_guidance}

PERFORMANCE WEIGHTS
- Exploration weight: {weights.get('exploration_weight', 1.0):.1f}
- Efficiency weight: {weights.get('efficiency_weight', 1.0):.1f}
- Backtrack threshold: {weights.get('backtrack_threshold', 0.7):.1f}
- Dead end confidence: {weights.get('dead_end_confidence', 0.8):.1f}
\end{lstlisting}

\subsection{Orchestration Agent Prompt Design}\label{appendix:orchestrator_agent_instructions}
\paragraph{A. Model System Prompt)}
\begin{lstlisting}[style=prompt]
You are the Maze Strategy Orchestrator with REAL-TIME DECISION AWARENESS.

COORDINATE SYSTEM
- Grid maze: 'W' (Wall), 'O' (Open), 'E' (Exit).
- MATRIX coordinates (row, col); NORTH=-row, SOUTH=+row, EAST=+col, WEST=-col.

YOUR CAPABILITIES
1. Real-time decision contexts per agent (positions, scores, weights, unexplored dirs).
2. Movement conflicts (local penalties vs global exploration value).
3. Coordination opportunities (overlap/duplication).
4. Global optimization patterns (bottlenecks, gaps).

STRATEGIC RESPONSIBILITIES
1. Validate dead ends: flag incorrect markings against discovered cells.
2. Resolve movement conflicts: where efficiency penalties block global exploration.
3. Coordinate agents: divide unexplored areas to maximize coverage.
4. Break local minima: recommend overrides or temporary weight relaxations.
5. Keep guidance decision-aware: amplify agents' local context, do not blindly overwrite.

RESPONSE CONTRACT (STRICT)
- Output a SINGLE JSON object (no prose, no code fences).
- Keys: "analysis", "corrections", "guidance_for_agents".
- corrections.remove_dead_ends: list of [row, col].
- corrections.add_exploration_focus: list of [row, col].
- guidance_for_agents: mapping agent_id -> short, actionable directive.
- Be specific but concise. Avoid chain-of-thought; summaries only.

VALIDATION GUARDRAILS
- If uncertain, return empty lists/objects.
- Never invent agent_ids or coordinates not in context.
- JSON must be valid UTF-8, no trailing commas, no comments.
\end{lstlisting}

\paragraph{B. System Context Message (runtime).}
\begin{lstlisting}[style=prompt]
Task: Find maze exit. Current Maze Exploration Analysis:

Orchestration Data (JSON): {{orchestration_data_json}}

ENHANCED DECISION INTELLIGENCE
- Movement Conflicts: {{movement_conflicts_json}}
- Exploration Coordination: {{exploration_coordination_json}}
- Efficiency Optimization: {{efficiency_optimization_json}}

FOCUS AREAS
- Dead end validation accuracy: {{dead_end_analysis_json}}
- Agent coordination summaries: {{agent_summaries_json}}
- Exploration coverage: {{discovered_cells_count}} cells
- Real-time decision contexts available: {{num_agents_with_context}} agents

INSTRUCTIONS
- Use the above decision data to produce ONLY the JSON object
  defined in the response contract. No markdown, no extra text.
\end{lstlisting}

\paragraph{C. Response Contract (sent to execution agent node)}
\begin{lstlisting}[style=prompt]
{
  "analysis": "Short decision-aware synthesis of conflicts/coordination and key gaps.",
  "corrections": {
    "remove_dead_ends": [ [{{r1}}, {{c1}}], [{{r2}}, {{c2}}] ],
    "add_exploration_focus": [ [{{r3}}, {{c3}}], [{{r4}}, {{c4}}] ]
  },
  "guidance_for_agents": {
    "{{agent_id_0}}": "Actionable, context-grounded instruction (e.g., override penalty and go EAST).",
    "{{agent_id_1}}": "Actionable instruction tailored to their movement scores/unexplored dirs."
  }
}
\end{lstlisting}

\subsection{Maze Generation and Complexity Metrics}\label{appendix:maze_generation_procedure}

\subsubsection{Maze Generation Algorithm}

Our experimental evaluation employs a custom maze generation framework derived from the AMaze benchmark \citep{godin2025amaze}, enhanced with Shannon entropy-based complexity measures to create systematic long-horizon task environments. The maze generation algorithm combines recursive backtracking with entropy-guided optimization to produce structured sequential decision-making challenges suitable for evaluating multi-agent coordination in extended task horizons.

\paragraph{Core Generation Process}

The maze generation follows a modified recursive backtracking algorithm that operates on a discrete grid $G \in \{W, O, E, X\}^{n \times n}$, where $W$ represents walls, $O$ denotes open paths, $E$ indicates the exit position, and $X$ marks the outer boundary frame. The algorithm prioritizes creating environments with extended solution sequences and multiple decision points:

\begin{itemize}[nosep]
    \item \textbf{Distributed Initialization:} Multiple starting points $S = \{s_1, s_2, \ldots, s_j\}$  are strategically placed across the maze to ensure complex path structures requiring sustained exploration, with $j$ varying based on maze size according to:
    \[
        j = 
        \begin{cases}
        1 & \text{if } n < 15 \\
        5 & \text{if } 15 \leq n < 25 \\
        9 & \text{if } n \geq 25
        \end{cases}
    \]

    \item \textbf{Entropy-Enhanced Carving:} From each starting point, the algorithm applies recursive backtracking with a dead-end factor $\delta \in [0.03, 0.35]$ that controls the density of decision points and backtracking requirements. Path carving continues until connectivity requirements create sufficiently long action sequences.

    \item \textbf{Quality Validation:} Generated mazes undergo multi-criteria validation including connectivity ratio $\rho \in [0.10, 0.95]$, minimum path length requirements, and Shannon complexity thresholds to ensure extended task horizons and multiple sequential decision points.
\end{itemize}

\subsubsection{Exit Placement Optimization}

Exit positions are optimized using a multi-objective scoring function that maximizes task horizon length and decision complexity:
\[
\text{Score}(p) = 10 \cdot d_{\text{path}}(s, p) + 5 \cdot d_{\text{Manhattan}}(s, p) + \phi_{\text{edge}}(p) + \phi_{\text{topology}}(p) + 2 \cdot d_{\text{Manhattan}}(p, c)
\]
where:
\begin{itemize}[nosep]
    \item $d_{\text{path}}(s, p)$ is the shortest path distance from start $s$ to position $p$
    \item $d_{\text{Manhattan}}(s, p) = |s_x - p_x| + |s_y - p_y|$
    \item $\phi_{\text{edge}}(p)$ rewards edge proximity (15 for edges, +25 for corners)
    \item $\phi_{\text{topology}}(p)$ rewards dead ends (+30) and penalizes junctions (-10)
    \item $c$ is the maze center
\end{itemize}

\subsubsection{Shannon Entropy-Based Complexity Measures}

Following the AMaze framework, we implement entropy-based metrics to quantify long-horizon task difficulty.

\paragraph{Surprisingness Metric}

The surprisingness $S(M)$ quantifies the entropy of directional decisions along the optimal path:
\[
S(M) = -\sum_{i \in \{W,O,E,X\}} p(i) \log_2 p(i)
\] where $p(i)$ is the empirical frequency of direction $i$ in the optimal trajectory. Higher values indicate greater planning unpredictability.

\paragraph{Deceptiveness Metric}

The deceptiveness $D(M)$ captures the entropy of trap transitions that lead to suboptimal paths:
\[
D(M) = \sum_{c \in C} \sum_{s \in T} -p(s|c) \log_2 p(s|c)
\]
where $C$ are cells adjacent to the optimal path, $T$ are trap states, and $p(s|c)$ is the transition probability from $c$ to $s$.

\paragraph{Trap Detection and Quantification}

Traps are defined as extended dead-end paths that test long-horizon strategy recovery. Each trap $t$ is characterized by:
\begin{itemize}[nosep]
    \item \textbf{Depth:} Length from branch to dead end
    
    \item \textbf{Branching Factor:} Number of branches within the trap
    
    \item \textbf{Weight:} 
    \[
    w_t = 1.0 + 0.5 \cdot \text{depth} + 0.3 \cdot \text{branches} + 0.2 \cdot \text{dead\_ends}
    \]
\end{itemize}

Total trap complexity is:
\[
T_c = \sum_{t \in \text{Traps}} w_t
\]

\subsubsection{Experimental Maze Collection}

We generated 15 mazes across four pre-set difficulty categories. We implemented easy difficulty as baseline only, and omitted the very hard difficulty for the analysis of this paper. The following setup was then used for experiments.

\begin{itemize}[nosep]
    \item \textbf{5x Easy:} $[10, 30]$ difficulty $ = 0.03$, size $12 \times 12$
    \item \textbf{5x Medium:} $[30, 60]$ difficulty $ = 0.10$, size $18 \times 18$
    \item \textbf{5x Hard:} $[60, 80]$ difficulty $ = 0.25$, size $25 \times 25$
    \item \textbf{0x Very Hard:} $[80, 95]$ difficulty $ = 0.35$, size $30 \times 30$
\end{itemize}

\noindent All maze specifications, including topology, complexity scores, and optimal paths, are available in the supplementary repository.


\end{document}